\documentclass[%
	aip,
	jmp,%
	amsmath,amssymb,
	reprint,%
	superscriptaddress
]{revtex4-1}

\usepackage{amsmath}
\usepackage{bm}         
\usepackage{color,soul}  
\usepackage{dcolumn}    
\usepackage{graphicx}   
\usepackage{multirow}

\newcommand{\quotes}[1]{``#1''}
\newcommand{\bra}[1]{\ensuremath{\langle \: #1 \: |}}
\newcommand{\ket}[1]{\ensuremath{| \: #1 \: \rangle}}
\newcommand{\OvI}[2]{\ensuremath{\langle \: #1 \mid #2 \: \rangle}}
\newcommand{\ExV}[3]{\ensuremath{\langle \: #1 \mid #2 \mid #3 \: \rangle}}
\newcommand{\w}{\ensuremath{\omega}}

\usepackage{titlesec}
\titlespacing*{\subsection}
	{0pt}{2.5ex}{0.5ex}
\titlespacing*{\section}
	{0pt}{3.5ex}{1.0ex}

\begin{document}
	
\title{Optimization of highly excited matrix product states with an application to vibrational spectroscopy}

\author{Alberto Baiardi}
\affiliation{ 
	ETH Z\"urich, Laboratorium f\"ur Physikalische Chemie, Vladimir-Prelog-Weg 2, 8093 Z\"urich, Switzerland
}
\author{Christopher J. Stein}
\affiliation{ 
	ETH Z\"urich, Laboratorium f\"ur Physikalische Chemie, Vladimir-Prelog-Weg 2, 8093 Z\"urich, Switzerland
}
\author{Vincenzo Barone}
\email[Corresponding author: ]{vincenzo.barone@sns.it}
\affiliation{ 
	Scuola Normale Superiore, Piazza dei Cavalieri 7, 56126 Pisa, Italy
}
\author{Markus Reiher}
\email[Corresponding author: ]{markus.reiher@phys.chem.ethz.ch}
\affiliation{ 
	ETH Z\"urich, Laboratorium f\"ur Physikalische Chemie, Vladimir-Prelog-Weg 2, 8093 Z\"urich, Switzerland
}

\date{\today}

\begin{abstract}
Configuration-interaction-type calculations on electronic and vibrational structure are often the method of choice for the reliable approximation of many-particle wave functions and energies.
The exponential scaling, however, limits their application range. In vibrational spectroscopy, for example, molecules with more than 15 to 20 vibrational modes can hardly be studied.
An efficient approximation to the full configuration interaction solution can be obtained with the density matrix renormalization group (DMRG) algorithm without a restriction to a predefined excitation level.
In a standard DMRG implementation, however, excited states are calculated with a ground-state optimization in the space orthogonal to all lower lying wave function solutions.
A trivial parallelization is therefore not possible and the calculation of highly excited states becomes prohibitively expensive, especially in regions with a high density of states.
Here, we introduce two variants of the density matrix renormalization group algorithm that allow us to target directly specific energy regions and therefore highly excited states. The first one, based on shift-and-invert techniques, is particularly efficient for low-lying states, but is not stable in regions with a high density of states. The second one, based on the folded auxiliary operator, is less efficient, but more accurate in targeting high-energy states.
We apply the algorithm to the solution of the nuclear Schr\"{o}dinger equation, but emphasize that it can be applied to the diagonalization of general Hamiltonians as well, such as the electronic Coulomb Hamiltonian to address X-ray spectra.
In combination with several root-homing algorithms and a stochastic sampling of the determinant space, excited states of interest can be adequately tracked and analyzed during the optimization.
We validate these algorithms by calculating several highly excited vibrational states of ethylene and demonstrate that we can accurately calculate prominent spectral features of large molecules such as the sarcosine-glycine dipeptide.

\end{abstract}

\maketitle

\setlength{\parindent}{0cm}
\setlength{\parskip}{0.6em plus0.2em minus0.1em}

\section{Introduction}

In the Born-Oppenheimer approximation, molecular vibrations are defined by the electronic potential in which the nuclei are moving.
This potential is usually anharmonic and requires correlated methods to account for the strong coupling between vibrational modes.
Many methods have been developed over the years to face the necessity of including a large number of configurations in the final wave function and capture a sufficient part of the correlation energy.
Quite frequently, these methods were adapted from electronic structure theory, such as configuration interaction (CI),\cite{Carter1997_VCI,Bowman2003_Multimode,Carrington2008_EnergyLevelsCH5,Neff2009_LargeScaleVCI,Scribano2010_VCIFast,Jacob2014_VCI-LocalModes,Carrington2016_RotationVibration-CH5+,Carrington2017_Review} perturbation theory approaches\cite{Sibert1988_VV-PT,Christiansen2003_VMP2,Barone2005_VPT2,Krasnoshchekov2012_CVV-PT,Bloino2012_Properties,Stanton2018_VPT4} and coupled-cluster expressions.\cite{Christiansen2004_VCC,Prasad2008_VCC-BosonicRepresentation,Seidler2009_VCCAutomatic}
Due to their exponential scaling, the full-CI approach, in which all configurations in a given basis set are included in the wave function expansion, is in most cases truncated to certain excitation ranks in order to limit the number of configurations considered.
Truncated CI is, however, strongly dependent on the choice of the basis functions.
Very recently, we adapted the density matrix renormalization group (DMRG)\cite{whit92,whit93,scho11} algorithm to the optimization of vibrational wave functions (vDMRG).\cite{Baiardi2017_vDMRG}
vDMRG combines two strategies that are largely employed in large-scale VCI calculations, \textit{i.e.} precontraction\cite{Meyer2004_Precontraction,Carrington2017_Precontraction} and pruning techniques.\cite{Bowman1991_BasisCoupling,Handy2004_MultimodeBasisPruning,Rauhut2007_ConfigurationSelection,Pouchan2010_VCI-P,Carrington2012_Pruning,Bowman2015_BasisPruningCH3NO2,Carrington2017_PrunedBasis} Like the former methods, vDMRG energies are obtained from the representation of the vibrational Hamiltonian in a precontracted basis, obtained as linear transformation of the full-dimensional one. vDMRG optimize iteratively this precontrated basis, unlike the majority of alternative VCI approaches, where the contraction is determined once and kept constant in the simulation. In vDMRG, the dimension of this precontracted basis would grow at each step of the optimization. To avoid this increase in the dimension of the basis, it is truncated at each optimization step by including only the configurations, giving the best representation of the wave function in a least-squares sense. From this perspective, vDMRG can be interpreted as an iterative pruning scheme.

Like vDMRG, the rank reduced block power (RRBP) method reduces the computational cost of VCI through a tensor factorization of the CI matrix.\cite{Leclerc2017,Carrington2018_CP-RRPM} However, DMRG optimizes a matrix product state (MPS) parametrization, while RRBP expresses the wave function as a canonical polyadic (CP) tensor. Unlike MPSs, the CP format is not designed to reproduce strong correlation between modes. For this reason, the dimension of the CP tensors required to obtain converged vibrational energies is much higher than for DMRG. This might also be the reason why the CP format has not been applied to electronic-structure problems, while DMRG has become a reference method for strongly correlated systems.

The calculation of excited states with DMRG represents a challenge both from a fundamental and an algorithmic perspective. First, it is still not clear to what extent the MPS parametrization can efficiently encode excited-state wave functions. Furthermore, the DMRG optimization algorithm is designed for ground states and must be generalized to target excited states efficiently. This generalization is most challenging in high-lying dense regions of the spectrum of a Hamiltonian. Prominent examples are found in X-ray absorption and vibrational spectroscopies. In our initial implementation of vDMRG,\cite{Baiardi2017_vDMRG} excited states were optimized with a standard DMRG ground-state search in the space orthogonal to the already optimized wave functions of the lower-lying vibrational states.
Therefore, all lower-lying wave functions need to be calculated and sufficiently converged in order to optimize a given excited state.\cite{McCulloch2007_FromMPTtoDMRG,Keller2015_MPS_MPO}
This is a major drawback, especially for highly excited vibrational states of large molecules, because the sequential optimization of all states starting from the vibrational ground state cannot be trivially parallelized. Furthermore, in regions with a high density of states, the convergence is often very slow due to root flipping events.

To overcome such limitations, more refined diagonalization schemes were devised for traditional approaches to allow for the calculation of excited states in both electronic and vibrational structure problems.
For electronic structure problems, we may refer to the energy-specific Davidson approach, that has been successfully employed for the calculation of excitation energies\cite{Li2011_EnergySpecific} and ionization potentials\cite{Lestrange2015_EnergySpecificXAS} with time-dependent density functional\cite{Stratmann1998_DavidsonTDDFT} and equation of motion coupled-cluster theories.\cite{Li2015_XASEOMCC}
Applications of energy-specific algorithms to vibrational structure problems have recently been introduced both for standard vibrational configuration interaction (VCI) algorithms\cite{Csaszar2009_VariationalLanczos,Rauhut2017_InteriorEigenvalues} and for vibrational wave functions expressed in tensor train (TT) format.\cite{Oseledets2016_TTVDMRG}
In the energy-specific Davidson algorithm, the eigenvectors are approximated by a Krylov-subspace iterative process, where only eigenvectors with energies above a certain threshold are kept in the update of the space.

In this work, targeting of pre-selected vibrational levels is achieved by mapping the original Hamiltonian $\mathcal{H}_\text{vib}$ onto an auxiliary operator $\Omega_\omega$, whose ground state corresponds to one of the interior eigenfunctions of $\mathcal{H}_\text{vib}$.\cite{Fokkema1998_JacobiDavidsonQR,Sleijpen2000_JacobiDavidson,Ventra2002_TargetedDiagonalization}
By applying standard iterative methods to the auxiliary operator, it is therefore possible to optimize the interior eigenfunctions of $\mathcal{H}_\text{vib}$. Out of the different functional forms for $\Omega_\omega$ that have been proposed in the literature,\cite{Leforestier1995_SpectralTransformation,Carrington2001_SpectralTransformation,Csaszar2009_VariationalLanczos} we will employ the shift-and-invert (S\&I)\cite{Ruhe1980_SandI,Ventra2002_TargetedDiagonalization} and the folded\cite{VanVoorhis2017_SigmaSCF} operators. The main advantage of the former is the possibility of exploiting the Harmonic Ritz Values theory\cite{Goossens1999_HarmonicRitzValues} to avoid the explicit inversion of the Hamiltonian. However, the S\&I method can only be applied to local diagonalization problems, but not to the full operator encoded as matrix product operator (MPO). On the contrary, the folded operator method can be easily extended to full MPOs with a significant increase in the reliability of the algorithm. Current, state-of-the art energy-specific DMRG implementations are based on S\&I auxiliary operators only.\cite{Dorando2007_TargetingExcitedStates,Yu2017_ShiftAndInvertMPS} However, in this work we show that, in order to target highly-excited states, building the auxiliary operator from the local representation of the Hamiltonian is not sufficient to achieve a fast and robust convergence. In such cases, the folded spectrum approach, which corresponds to a well-defined variational principle, ensures convergence, however at the price of higher computational cost due to the need of encoding the squared Hamiltonian as an MPO.

The robustness of the previous algorithms can be further increased by combination with a root-homing algorithm,\cite{Kammer1976_RootHoming} which has already been employed in both electronic\cite{Kovyrshin2010_OptimizationStateSelective,Kovyrshin2011_SelectiveTDDFT} and vibrational\cite{Reiher2003_ModeTrackingNanotubes,Reiher2004_ModeTracking,Reiher2007_ReviewModeTracking,lube09} problems to consistently follow the correct root during the optimization. A maximum-overlap criterion, which is equivalent to root-homing, has recently been proposed to optimize many-body localized states with DMRG for spin chains.\cite{Khemani2016_ExcitedStateDMRGSpatial,Devakul2017_DMRGXX} The resulting approach, known as DMRG-X, has, however, not been combined with iterative diagonalization schemes, therefore preventing its application to large systems.

We note that time-dependent (TD) DMRG,\cite{Ronca2017_TDDMRG-Targeting,Ma2017,Yao2018_AdaptiveTDDMRG} where the time-dependent Schr\"{o}dinger equation is solved explicitly and excited-state properties are extracted through Fourier transformation of an appropriate autocorrelation function, is an alternative method to the one proposed in this paper. The main difference between TD-DMRG and energy-specific formulation introduced here is that the former avoids the explicit calculation of eigenfunctions. This can be a major limitation if, for example, a perturbative correction, which requires the eigenfunctions, must be evaluated after DMRG optimization.

This paper is organized as follows. In the first part of Sec.~\ref{sec:theory}, the details of the energy-specific DMRG algorithms are presented, and their implementation within vDMRG is described in detail. 
Then, different maximum-overlap variants of vDMRG, based on root-homing algorithms, are defined. 
Finally, a stochastic CI coefficient sampling method is discussed and compared to its electronic structure counterpart.\cite{Moritz2007_CIReconstruction,Boguslawski2011_SRCAS} 
After a brief overview of the details of the implementation in Sec.~\ref{sec:details}, the energy-specific variants of vDMRG are applied in the calculation of highly excited states of ethylene and the sarcosine-glycine dipeptide, whose low-energy states have already been studied with the standard variant of vDMRG.\cite{Baiardi2017_vDMRG}

\section{General theoretical framework}
\label{sec:theory}

The presentation of the theoretical section is divided in three parts. First, we revise the standard formulation of vDMRG. Then, we discuss the energy-specific formulations of vDMRG, possibly coupled with root-homing to optimize predetermined vibrational levels. Finally, we extend an algorithm, originally devised for electronic wavefunctions to reconstruct the CI form of a wavefunction encoded as MPS, to vibrational wavefunctions. Withing the first two parts, we will assume a general form for the Hamiltonian $\mathcal{H}$ and, as a consequence, the theory applies to both vibrational and electronic problems. The theory presented in the final part applies to vibrational Hamiltonians $\mathcal{H}_\text{vib}$ only.

\subsection{Energy-specific DMRG}

Before describing the details of the S\&I algorithms, we recall some basic properties of the DMRG-optimized wave function and operators.

In DMRG, a wave function $\ket{\Psi}$ for an $L$-body system can be expressed as a matrix product state (MPS)\cite{Rommer1997_MPS} as follows:
\begin{equation}
  \ket{\Psi^{(k)}} = \sum_{\sigma_1,...,\sigma_L}^{N_\text{max}} \sum_{a_1,...,a_\text{L-1}}^m
  {M}_{1,a_1}^{\sigma_1} {M}_{a_1,a_2}^{\sigma_2} \cdot
  {M}_{a_\text{L-1},1}^{\sigma_L}
  \ket{\sigma_1,...,\sigma_L} \, .        
  \label{eq:DMRG_ansatz}
\end{equation}
The basis states are occupation number vectors (ONVs) $\ket{\sigma_1,...,\sigma_L}=\ket{\boldsymbol{\sigma}}$ where each local basis has dimension $N_\text{max}$ and the $\textbf{M}^{\sigma_i} = \{ M_{a_\text{i-1},a_\text{i}}^{\sigma_i} \}$ are site matrices of maximum dimension $m \times m$ (note that $\mathbf{M}^{\sigma_1}$ and $\mathbf{M}^{\sigma_L}$ are row and column vectors, respectively), where $m$ is the number of renormalized block states (also called bond dimension) and a site denotes the position of a single particle basis (orbital or vibrational mode) on the DMRG lattice.
We note that the MPS structure of the wave function is equivalent to the TT format.\cite{osel11,Oseledets2016_TTVDMRG}
Although the energy is a non-linear function of these matrix entries, the variational optimization is efficiently carried out by a sequential iterative optimization of $\textbf{M}^{(k)\sigma_l}$ for each site $l$, starting from $l=1$ and going back and forth along the one-dimensional lattice of sites, a process which is referred to as \quotes{sweeping}.
The sequence of $L$ site optimizations is referred to as one macroiteration step in the sweep algorithm (or one \quotes{sweep}), whereas the optimization of an individual site is called a microiteration step. In the case of vibrational wave functions, $L$ may be chosen to either represent vibrational modes (as chosen here) or as a label for all ground- and excited-state basis functions of all modes.

We emphasize that our implementation is flexible with respect to the choice of the Hamiltonian (electronic or vibrational) that may be specified on input. A Hamiltonian $\mathcal{H}$, expressed as an MPO, reads,
\begin{equation}
	\mathcal{H} = \sum_{\bm{\sigma\sigma'}}^{N_\text{max}} \sum_{b_1,\dots,b_{L-1}}^{b_\text{max}}
	  W_{1 b_1}^{\sigma_1,\sigma_1'} \cdots W_{b_{l-1} b_l}^{\sigma_l,\sigma_l'} \cdots W_{b_{L-1} 1}^{\sigma_L,\sigma_L'} \ket{\bm{\sigma}} \bra{\bm{\sigma'}}\, .
	\label{eq:mpo_general}
\end{equation}

Here, $\bm{W}^{\sigma_l,\sigma_l'} = \{ W_{b_{l-1},b_l}^{\sigma_l,\sigma_l'} \}$ collects all coefficients of the matrix representation of the Hamiltonian, which consist of strings of ladder operators, acting on site $l$. In this work, sites are represented in the harmonic oscillator basis. However, neither the theoretical foundations nor the implementation are restricted to this special choice of site functions. In the single-site version of DMRG, the energy is minimized with respect to the tensor associated to a single site $l$, by keeping all the other tensors fixed. The minimization leads to the following eigenvalue equation:
\begin{equation}
 \begin{aligned}
  \sum_{\sigma_l'}^{N_\text{max}} \sum_{a_{l-1}'a_l'}^m \sum_{b_{l-1},b_l}^{b_\text{max}}
  W_{b_{l-1}b_l}^{\sigma_l\sigma_l'} L_{a_{l-1}a_{l-1}'}^{b_{l-1}} & M_{a_{l-1}'a_l'}^{\sigma'_l} R_{a_l'a_l}^{b_l} \\
  =& E M_{a_{l-1}a_l}^{\sigma_l} \, ,
 \end{aligned}
 \label{eq:DMRG_eigenvalue}
\end{equation}
where the tensors $\bm{L} = L_{a_l,a_l'}^{b_l}$ and $\bm{R} = R_{a_l,a_l'}^{b_l}$ are obtained through MPS-MPO contractions of the sites to the left and right of site $l$, respectively, and $E$ is the energy of the state. 

After its optimization, the tensor $\bm{M}^{\sigma_l}$ is reshaped as a $N_\text{max} m \times m$ matrix, where $N_\text{max}$ is the maximum number of basis states (e.g. the number of possible occupation numbers of a spatial orbital or of harmonic oscillator basis function) per site $l$, and orthogonalized by singular value decomposition.
Before this decimation, $\bm{M}^{\sigma_l}$ is a $N_\text{max} m \times m$ matrix, where $N_\text{max}$ is the maximum number of basis states (e.g. the number of possible occupation numbers of a spatial orbital or of harmonic oscillator basis function) per site $l$. 
After decimation, the dimension of $\bm{M}^{\sigma_l}$ is reduced to $m \times m$.

In the standard variant of DMRG, the ground state energy is optimized variationally and Eq.~(\ref{eq:DMRG_eigenvalue}) is solved with an iterative eigensolver that targets one end of the eigenvalue spectrum, such as the Davidson\cite{Davidson1975_DavidsonDiagonalization} and Jacobi-Davidson\cite{Sleijpen2000_JacobiDavidson} algorithms. The subsequent optimization of all vibrational states can be accomplished by the Liu-Davidson algorithm \cite{Kosugi1984_LiuDavidson} which is a generalization of the Davidson method for the simultaneous calculation of several eigenpairs.
These eigenpairs are optimized in each iteration, while the vector space is enlarged through the application of the standard Davidson method for each unconverged root.
Despite its simplicity, this approach has several drawbacks. 
First of all, the cost is severely increased compared to the standard Davidson approach, especially when a large number of eigenstates is calculated. Moreover, the Liu-Davidson algorithm can be applied to calculate several eigenpairs of the same operator. However, in DMRG, the boundaries $\bm{L}$ and $\bm{R}$ depend on the targeted vibrational state and, hence, a different operator appears in Eq.~(3) for each state. We highlight that, in state-averaged formulations of DMRG, the boundaries of each state are averaged to obtain a common set of boundaries for all the targeted states. However, as discussed in Ref. 38, the convergence of state-average DMRG formulations is significantly slower than that of state-specific DMRG, and therefore they will not be discussed in the present paper.

As mentioned in the introduction, we employ an S\&I algorithm to overcome the problems described above.
We define an auxiliary operator $\Omega_\omega$ (referred to as S\&I operator in the following), whose representation in a given basis set is,
\begin{equation}
  \bm{\Omega}_\omega = \left( \omega \bm{I} - \bm{H} \right)^{-1} = \bm{H}_\omega^{-1} \, ,
  \label{eq:Omega_operator_def}
\end{equation}
where $\omega$ is an energy shift and $\bm{H}$ is the representation of the Hamiltonian in the same basis set. As will be discussed below, the choice of this basis strongly affects the efficiency of DMRG[S\&I]. If not otherwise specified, the representation is built from the renormalized basis for site $l$,\cite{scho11,Keller2015_MPS_MPO}
\begin{equation}
  H_{ (a_{l-1} \sigma_l a_l , a_{l-1}' \sigma_l' a_l') } = 
      \langle a_{l-1} \sigma_l a_l | \mathcal{H} | a_{l-1}' \sigma_l' a_l' \rangle \, ,
  \label{eq:HVib_representation}
\end{equation}
where $l$ is the index of the site which is optimized. The left and right renormalized bases ($| a_{l-1} \rangle$ and $| a_l \rangle$) are obtained by contracting the tensors before and after the $l$-th site; their definition can be found, for example, in Ref.~\citenum{Keller2015_MPS_MPO}.
The renormalized basis spans only a small subset of the full Hilbert space.

The smallest eigenvalue of $\Omega_\omega$ corresponds to the first eigenvalue of $\bm{H}$ larger than $\omega$. 
Hence, this interior eigenvalue of $\bm{H}$ can be accessed by applying iterative eigensolvers designed to target eigenpairs at one end of the energy spectrum to $\Omega_\omega$. 
In the following, we associate to a tensor $M_{a_{l-1}a_l}^{\sigma_l}$ a vector $| \nu \rangle$ belonging to the product basis $\ket{a_{l-1} \sigma_l a_l}$ defined as follows:

\begin{equation*}
	\ket{\nu} = \sum_{a_{l-1}=1}^m \sum_{a_l=1}^m \sum_{\sigma_l=1}^{N_\text{max}} M_{a_{l-1}a_l}^{\sigma_i} 
	\ket{a_{l-1} \sigma_l a_l}
	\label{eq:RenormalizedBasis_Expansion}
\end{equation*}

In the Davidson algorithm, $\ket{\nu}$ is expanded in a subspace $\ket{\boldsymbol{\eta}} = \left( \ket{\eta_1}, ... , \ket{\eta_n} \right)$ (we refer to this space as the search space) of the full vector space as

\begin{equation}
  \ket{\nu_n} = \sum_{i=1}^n c_i^{(n)} \ket{\eta_i} \, ,
  \label{eq:Subspace_expansion}
\end{equation}
where $n$ specifies the total number of iterations in the Davidson algorithm.

The explicit matrix inversion of Eq.~(\ref{eq:Omega_operator_def}) can be avoided with the Harmonic Ritz values theory.\cite{Goossens1999_HarmonicRitzValues} In standard Davidson diagonalization, $\left(| \eta_1 \rangle , \ldots, |\eta_n \rangle \right)$ represents both the search space and the vector space in which the eigenvalue problem is solved (called test space in the following). 
The inversion of the matrix $\bm{H}_\omega$ can be avoided if the search space and the test space are different vector spaces.
Keeping $\left(| \eta_1 \rangle , \ldots, |\eta_n \rangle \right)$ as the search space, we define the test space $ \ket{ \tilde{\boldsymbol{\eta}}}$ as
\begin{equation}
 \ket{ \tilde{\eta}_k } = \sum_{l=1}^n \left( \bm{H}_\omega \right)_{kl} \ket{\eta_l} \, .
  \label{eq:Reformulate}
\end{equation}

As discussed in Refs.~\citenum{Sleijpen2000_JacobiDavidson} and \citenum{Paige1995_HarmonicDavidson}, this procedure, known as oblique projection, leads to the following generalized eigenvalue problem:

\begin{equation}
  \sum_{l=1}^n \bra{\eta_k} \bm{H}_\omega \ket{\eta_l} \, \left( \textbf{c}^{(n)} \right)_l = 
  \frac{1}{\text{E}_{\omega,1}^{(n)}} \sum_{l=1}^n \langle \tilde{\eta}_k | \tilde{\eta}_l \rangle \, \left( \textbf{c}^{(n)} \right)_l \, .
  \label{eq:Reformulate_3}
\end{equation}

where $\bm{c}^{(n)}$ collects the linear coefficients of the expansion of the eigenvectors in the subspace. In this way, the lowest eigenvalue of $\Omega_\omega$, $\left(\text{E}_{\omega,1}^{(n)}\right)^{-1}$, is calculated without explicitly inverting the MPO associated to the Hamiltonian.

The $\left(| \tilde{\eta}_1 \rangle , \ldots, \ket{\tilde{\eta_n}} \right)$ basis is, in general, not orthogonal and $\OvI{\tilde{\boldsymbol{\eta}}}{ \tilde{\boldsymbol{\eta}}} \neq \mathcal{I}$. Hence, Eq.~(\ref{eq:Reformulate_3}) is a generalized eigenvalue problem. Following an approach already introduced in the context of DMRG,\cite{Dorando2007_TargetingExcitedStates} the $\left(| \tilde{\eta}_1 \rangle , \ldots, \ket{\tilde{\eta_n}} \right)$ basis set can be orthogonalized through a Gram-Schmidt algorithm,
\begin{equation}
  \ket{\tilde{\eta}_i'} = \ket{\mathcal{H}_\omega \eta_i'} = 
                          \ket{\tilde{\eta}_i} - \sum_{j<i} 
  						  \frac{\OvI{\tilde{\eta}_i}{\tilde{\eta}_j'}}
       						   {\OvI{\tilde{\eta}_j'}{\tilde{\eta}_j'}} 
       				      \ket{\tilde{\eta}_j'} \, .
  \label{eq:GSOrtho_EtaPrimed}
\end{equation}

This simplifies the eigenvalue problem of Eq.~(\ref{eq:Reformulate_3}), because $\OvI{\tilde{\boldsymbol{\eta}}}{ \tilde{\boldsymbol{\eta}}} = \mathcal{I}$. Consequently, the $\{ \ket{\eta_i'} \}$ basis, where $\ket{\eta_i'} = \bm{H}_\omega^{-1} \ket{ \tilde{\eta}_i' } $, must be updated as follows:
\begin{equation}
	\ket{\eta_i'} = \ket{\eta_i} - \sum_{j<i} 
					\frac{\OvI{\tilde{\eta}_i}{\tilde{\eta}_j'}}
	                     {\OvI{\tilde{\eta}_j'}{\tilde{\eta}_j'}} 
					\ket{\eta_j'} \, .
  \label{eq:GSOrtho_EtaPrimed_2}
\end{equation}

Eq.~(\ref{eq:Reformulate_3}) can now be expressed in the orthogonal basis $\ket{\tilde{\boldsymbol{\eta}}'}$,
\begin{equation}
 \sum_{l=1}^n \bra{\eta_k'} \bm{H}_\omega \ket{\eta_l'} \, \left( \tilde{\textbf{c}}^{(n)} \right)_l
 = \frac{1}{\text{E}_{\omega,1}^{(n)}} \left( \tilde{\textbf{c}}^{(n)} \right)_k \, ,
  \label{eq:Reformulate_4}
\end{equation}
which is an ordinary eigenvalue problem. 
As described in Ref.~\citenum{Dorando2007_TargetingExcitedStates} and in the supplementary material, the search space is expanded according to the standard Davidson algorithm.

The Jacobi-Davidson (JD)\cite{Olsen1990_Iterative,Sleijpen2000_JacobiDavidson} algorithms differ in the expansion step. 
If $\ket{\nu_n}$ is the $n$-th approximation to the lowest-energy eigenvector, the ($n+1$)-th is constructed from the following equation:
\begin{align}
  \left(  \bm{I} - \ket{\nu_\text{n}}\bra{\nu_\text{n}} \right)
  &\left( \bm{H} - \text{E}_1^\text{(n)} \bm{I} \right)\nonumber \\ 
 \times& \left( \bm{I} - \ket{\nu_\text{n}}\bra{ \nu_\text{n}}\right) 
  \ket{\eta_{n+1}}
  = -\ket{ r_\text{n}} \, ,
  \label{eq:JD_expansio}
\end{align}
with residual $\ket{r_n} = \left( \bm{H} - \mathcal{I} \text{E}_1^\text{(n)} \right) \ket{\nu_\text{n}}$ (a detailed discussion of the JD algorithm can be found in Ref.~\citenum{Sleijpen2000_JacobiDavidson}). 
If $\text{E}_1^\text{(n)}$ corresponds to the exact eigenvalue, $\ket{\eta_{n+1}}$ is the component of the exact eigenfunction in the space orthogonal to the search space. Hence, a direct inversion of Eq.~(\ref{eq:JD_expansio}) leads to convergence in a single iteration. However, $\text{E}_1^\text{(n)}$ is only an approximation of the exact eigenvalue, and hence, although Eq.~(\ref{eq:JD_expansio}) is solved exactly, the exact eigenfunction will still have non-zero components in the space orthogonal to the search space. For this reason, it is usually sufficient to solve Eq.~(\ref{eq:JD_expansio}) only approximately, with a few steps of an iterative solver, such as the generalized minimal residual (GMRES) algorithm.\cite{Saad1986_GMRES,Koch2000_Bratwurst}
Although the extension of the JD method to the calculation of Harmonic Ritz values is known in the literature,\cite{Sleijpen2000_JacobiDavidson} its application in quantum chemistry is scarce, especially in the context of DMRG, where mostly Harmonic Davidson methods are applied.\cite{Dorando2007_TargetingExcitedStates} 

Eq.~(\ref{eq:JD_expansio}) must be modified to support Harmonic Ritz values. A straightforward generalization would imply to replace $\bm{H}$ with $\Omega_{\omega}$. However, as for the Davidson case, the resulting equation would require the explicit inversion of $\bm{H}$. As discussed in the SI, the correction equation can be rearranged as follows:
\begin{align}
  \left( \bm{I} - \ket{\tilde{\nu}_\text{n}}\bra{ \tilde{\nu}_\text{n}} \right)
  \left( \bm{H}_\omega - \text{E}_{\omega,1}^\text{(n)} \bm{I} \right) \nonumber \qquad \qquad \qquad \\
 \times \left( \bm{I} - \ket{\nu_\text{n}} \bra{\tilde{\nu}_\text{n}} \bm{H}_\omega \right) 
  \ket{\eta_\text{n+1} } 
  = \ket{ \tilde{r}_\text{n}} \text{E}_{\omega,1}^\text{(n)} \, ,
  \label{eq:JD_correction_modified_2}
\end{align}
where $\ket{\tilde{\nu}_n}$ and $\ket{\tilde{r}_n}$ are the approximation of the eigenfunction and the related error at the $n$-th iteration expressed in the $\{ \nu_i' \}$ basis set. The correction equation is now independent of the inverse of $\bm{H}_\omega$. Eq.~(\ref{eq:JD_correction_modified_2}) can be solved as in standard Jacobi-Davidson problems, without any increase of the computational effort. 
The only additional step with respect to the standard Jacobi-Davidson algorithm is the Gram-Schmidt orthogonalization (Eqs.~(\ref{eq:GSOrtho_EtaPrimed}) and (\ref{eq:GSOrtho_EtaPrimed_2})).

As suggested in Ref.~\citenum{Sleijpen2000_JacobiDavidson}, an alternative form of the correction equation is:
\begin{equation}
	\begin{aligned}
     	 \left( \bm{I} - \frac{\ket{\nu_\text{n}}\bra{ \tilde{\nu}_\text{n}}}{\OvI{\tilde{\nu}_\text{n}}{ \nu_\text{n}}} \right)
	   & \left( \bm{H}_\omega - \text{E}_{\omega,1}^\text{(n)} \bm{I} \right) \\
 \times& \left( \bm{I} - \frac{\ket{\nu_\text{n}}\bra{ \tilde{\nu}_\text{n}}}{\OvI{\tilde{\nu}_\text{n}}{ \nu_\text{n}}} \right)
         \ket{\eta_\text{n+1} }   = -\ket{ r_\text{n}} \, ,
	\end{aligned}
  \label{eq:JD_correction_modified_3}
\end{equation}
which we will employ in this work because it can be easily combined with a deflation process (i.e., orthogonalization with respect to a converged root).\cite{Fokkema1998_JacobiDavidsonQR,Sleijpen2000_JacobiDavidson} The previous equation holds only if the bases $\{ \eta_i \}$ and $\{ \tilde{\eta}_i \}$ are biorthogonal. The two bases can be biorthogonalized following the algorithm described in Ref.~\citenum{Sleijpen2000_JacobiDavidson}.
We implemented the S\&I form of the Jacobi-Davidson diagonalization and applied to DMRG. The resulting algorithm will be referred to in the following as DMRG[S\&I].

\subsection{Folded DMRG}
\label{Sec:Folded}

In the previous section, $\bm{H}_\omega$ was the representation of $\mathcal{H}_\omega$ in the renormalized basis for a given microiteration step of a DMRG sweep. We will discuss now how the projection of $\mathcal{H}$ on this basis affects the algorithms described above. The representation of the Hamiltonian in the renormalized basis, $\bm{H}_\omega$, can be expressed as,
\begin{equation}
  \bm{H}_\omega = \mathcal{P}_l \mathcal{H}_\omega \mathcal{P}_l \, ,
  \label{eq:ProjectedHamiltonian}
\end{equation}
where $\mathcal{P}_l$ is the projection operator on the renormalized basis for the $l$-th site,
\begin{equation}
  \mathcal{P}_l = \sum_{a_{l-1}} \sum_{\sigma_l} \sum_{a_l} \ket{ a_{l-1} \sigma_l a_l } \bra{ a_{l-1} \sigma_l a_l } \, .
  \label{eq:ProjectionOperator}
\end{equation}
In the standard iterative diagonalization scheme (both Davidson and Jacobi-Davidson), the matrix elements to be computed ($\left(\bm{H}_{\omega}\right)_{kh}$) are of the following form:
\begin{equation}
  \left(\bm{H}_{\omega}\right)_{kh} = \langle \eta_k | \mathcal{P}_l \mathcal{H}_{\omega} \mathcal{P}_l | \eta_h \rangle
  \label{eq:MatrixElements_StandardDavidson}
\end{equation}
The two elements of the test space, $| \eta_k \rangle$ and $| \eta_h \rangle$, belong to the space spanned by the renormalized basis. For this reason, the action of the projection operator $\mathcal{P}_l$ does not alter these elements. As a consequence, $ \langle \eta_k | \mathcal{H}_{\omega} | \eta_h \rangle  = \langle \eta_k | \mathcal{P}_l \mathcal{H}_{\omega} \mathcal{P}_l | \eta_h \rangle$. Conversely, for the Harmonic Ritz value-based formulation of the iterative diagonalization schemes, also matrix elements of the following form,
\begin{equation}
  \langle \mathcal{H}_\omega \eta_k | \mathcal{H}_\omega \eta_h \rangle =
  \langle \eta_k | \mathcal{H}_\omega^2 | \eta_h \rangle \, ,
  \label{eq:MatrixElements_ModifiedDavidson}
\end{equation}
must be calculated. In this case, a different expression for Eq.~(\ref{eq:MatrixElements_ModifiedDavidson}) would be obtained employing the full Hamiltonian operator $\mathcal{H}_\omega$ instead of the projected one, $\bm{H}_\omega$. In fact, when $\bm{H}_\omega$ is applied twice to the vector $\ket{ \eta_h }$, the result of the first application of $\mathcal{H}_\omega$ on $\ket{ \eta_h }$ is implicitly projected onto the renormalized basis and then the Hamiltonian in applied on the resulting vector

\begin{equation}
	\left( \left( \bm{H}_\omega \right)^2 \right)_{kh} = \ExV{\eta_k}{\mathcal{H}_\omega \mathcal{P}_l \mathcal{H}_\omega}{\eta_h} \, .
	\label{eq:DoubleApplication}
\end{equation}

Conversely, if the full Hamiltonian is employed, the full Hamiltonian is applied directly to $\mathcal{H}_\omega$, without an intermediate projection. We recall that applying an operator to an MPS increases its rank. Therefore, the bond dimension of $\mathcal{H}_\omega | \eta_l \rangle$ is higher than the one of $| \eta_l \rangle$. On the contrary, $\bm{H}_\omega \ket{ \eta_h }$  has got the same bond size as $\ket{ \eta_h }$. Hence, the two vectors are different, and as a consequence

\begin{equation}
  \left( \left( \bm{H}_\omega^{(n)} \right)^2 \right)_{kl} \neq \ExV{\eta_k}{\mathcal{H}_\omega^2}{\eta_l} \, .
  \label{eq:Inequality}
\end{equation}

We have shown that applying the S\&I transformation to the local representation of the Hamiltonian implicitly introduces an approximation in the representation of the squared value of the Hamiltonian. We now assess the effect of this approximation on the accuracy of the energy-specific variants of DMRG. As already discussed for optimization algorithms of TT,\cite{osel11,Oseledets2016_TTVDMRG} the local eigenvalue problem given in Eq.~(\ref{eq:DMRG_eigenvalue}) is obtained from the minimization of the following functional:
\begin{equation}
  E[\ket{\Psi_\text{MPS}}] 
   = \min_{M^{\sigma_l}} \left\| \mathcal{H} | \Psi_\text{MPS} \rangle - E | \Psi_\text{MPS} \rangle \right\|^2
  \label{eq:ALS_minimization}
\end{equation}
where the minimization is with respect to the tensors $M^{\sigma_i}$ of rank $m$, where $m$ is the bond size of the MPS. As shown in Ref.~\citenum{Oseledets2016_TTVDMRG}, for positive-definite matrices the minimization of the functional given above is equivalent to the minimization of the following, simpler functional,
\begin{equation}
  E[\ket{\Psi_\text{MPS}}] = \min_{M^{\sigma_l}} \left ( \frac{ \langle \Psi_\text{MPS} | \mathcal{H} | \Psi_\text{MPS} \rangle }
				                                              { \langle \Psi_\text{MPS} | \Psi_\text{MPS} \rangle} \right)
  \label{eq:ALS_minimization_alternative}
\end{equation} 
However, even if the Hamiltonian is positive definite, its shift-and-invert counterpart $\Omega_\omega$ will have negative eigenvalues for a shift $\omega$ larger than the lowest eigenvalue. For this reason, the ALS minimization might not converge in this case. This issue has already been noted in the literature,\cite{Oseledets2016_TTVDMRG} but no cases in which the ALS minimization failed were detected. 

Here, we will employ a third, more robust DMRG formulation, where high-energy states are calculated as eigenvalues of the following auxiliary operator $\Omega_\omega^F$,
\begin{equation}
  \Omega_\omega^F = \left( \omega - \mathcal{H} \right)^2 \, ,
  \label{eq:FoldedOperator}
\end{equation}
usually referred to as folded operator. It is easy to show that the lowest eigenvalue of $\Omega_{\omega}^F$ is the eigenvalue of $\mathcal{H}$ which is closest to $\omega$. The folded functional has already been employed for targeting electronically excited states.\cite{neus16,VanVoorhis2017_SigmaSCF} Recently, a similar approach was studied in the context of DMRG\cite{Mach2013_FoldedTT} to calculate inner eigenvalues of operators expressed in TT format. The main advantage of $\Omega_\omega^F$ over $\Omega_\omega$ is that, in the first case, the spectral transformation is applied to the full Hamiltonian operator $\mathcal{H}$. The resulting, modified operator is only later projected in the renormalized basis and, therefore, an additional spectral transformation of its renormalized representation is not required. Its lowest energy eigenvalue (i.e., the eigenvalue with energy closer to the shift parameter $\omega$ employed in the spectral transformation) can instead be calculated with the standard, non-S\&I, Jacobi-Davidson algorithm. The matrix product operator representation of $\Omega_\omega^F$ can be obtained applying the same algorithm as for $\mathcal{H}$, following the procedure reported, for example, in Ref.~\citenum{scho11}. We note that the shift parameter is already included in the definition of the MPO. Hence, there is no need of a second shift of the local eigenvalue problem (Eq.~(\ref{eq:DMRG_eigenvalue})), and the following, standard correction equation,
\begin{equation}
	\begin{aligned}
		\left( \bm{I} - \ket{\nu_\text{n}}\bra{\nu_\text{n}} \right)
		      & \left( \bm{\Omega}_\omega^F - \left( \text{E}_{\omega,1}^\text{F(n)} \right)^2 \bm{I} \right) \\ 
        \times& \left( \bm{I} - \ket{\nu_\text{n}}\bra{ \nu_\text{n}}\right) 
		   \ket{\eta_{n+1}} = -\ket{ r_\text{n}} \, ,
	\end{aligned}
	\label{eq:CorrEq_Folded}
\end{equation}
can be directly evaluated in the Jacobi-Davidson algorithm. In the following, we will refer to this approach as folded DMRG (DMRG[f]). It has been already pointed out that the spectral range (i.e., the difference between the smallest and the largest eigenvalues) of the squared Hamiltonian is larger than the one of the original, non-squared Hamiltonian. This slows down the convergence of iterative diagonalization schemes, including the Jacobi-Davidson one. However, as we will discuss in Section IV, in regions with a high density of states DMRG[S\&I] is not stable due to the intrinsic approximation of the squared $\mathcal{H}^2$ operator. Even if slower, DMRG[f] ensures a much smoother convergence of the energy of the target state.

To conclude, once matrix elements of the squared Hamiltonian $\mathcal{H}^2$, defined as,
\begin{equation}
  \mathcal{H}^2 = \mathcal{H} \times \mathcal{H}
  \label{eq:SquaredHam_Def}
\end{equation}
are available, the variance, defined as,
\begin{equation}
  \sigma^2 = \langle \mathcal{H}^2 \rangle - \langle \mathcal{H} \rangle^2
  \label{eq:Variance}
\end{equation}
can be evaluated during the optimization. As already discussed in the literature,\cite{Devakul2017_DMRGXX,Hubig2018_ErrorEstimates} the variance is a reliable measure to probe the convergence of DMRG and must vanish when an MPS approaches the exact targeted eigenfunction. 

\subsection{Root-homing in vDRMG}
\label{Sec:root_homing}

In Refs.~\citenum{Dorando2007_TargetingExcitedStates} and \citenum{Yu2017_ShiftAndInvertMPS}, it was shown that the convergence of Davidson and Jacobi-Davidson diagonalization algorithms will be slow if multiple, almost degenerate excited states with an energy close to $\omega$ are present.
In fact, in regions with a high density of states, the lowest eigenpair of Eq.~(\ref{eq:Reformulate_4}), from which the Davidson or Jacobi-Davidson correction equation is built, might not correspond to the target state. This effect, also known as root flipping, lowers the efficiency of the optimization. To improve the convergence of the Jacobi-Davidson algorithm, here we propose an alternative algorithm, that exploits the locality of eigenstates on the DMRG lattice to improve the convergence of the diagonalization.
Our scheme is similar to the DMRG-X algorithm proposed recently,\cite{Khemani2016_ExcitedStateDMRGSpatial,Devakul2017_DMRGXX} where the root to be employed for the correction equation is not chosen only based on an energy criterion, but also on a locality one. A similar approach was also applied to the calculation of electronic excited states of molecular systems\cite{Dorando2007_TargetingExcitedStates,Chan2015_DMRG-GeomOpt} with a state-average approach (SA-DMRG).
Excited states that are close in energy are most often located on different sites of the DMRG lattice.
This allows us to consistently follow a single state during the optimization by selecting the eigenfunction that enters the correction equation with a \textit{root-homing} algorithm.\cite{Kammer1976_RootHoming}

Following an approach already proposed in the context of mode-tracking algorithms\cite{Reiher2003_ModeTrackingNanotubes,Reiher2004_ModeTracking} and more recently extended to the solution of Casida's equations in time-dependent density functional theory (TD-DFT),\cite{Kovyrshin2010_OptimizationStateSelective,Kovyrshin2011_SelectiveTDDFT} the correction equation can also be built from the eigenfunction with the largest overlap with a predefined element of the vector space (referred to in the following as \textit{test function}), and the resulting iterative diagonalization will converge the eigenfunction with the largest overlap associated with the test function.
We note that this approach is also equivalent to the maximum-overlap (MaxO) methods that are commonly applied for electronic structure problems.\cite{FloresMoreno2007,Gilbert2008,Baiardi2017_PropagatorVibronic}
If the optimized MPS deviates only slightly from the guess, the test function (denoted as \textit{test MPS} $\ket{\Phi_\text{test}}$  for the DMRG case) can be constructed by choosing the matrix entries in Eq.~(\ref{eq:DMRG_ansatz}) such that all ONVs but one vanish. In general, also an MPS obtained from a previous DMRG calculation, for example, with a lower value of $m$, can be used as test MPS.

Root-homing is independent on the form of the operator to diagonalize and, hence, can be coupled with both DMRG[f] and DMRG[S\&I]. We will denote the resulting algorithm as DMRG[f,MaxO] and DMRG[S\&I,MaxO], respectively.

We note that, in DMRG[MaxO], the $\omega$ parameter can be updated dynamically during the optimization.
In fact, as already mentioned in the previous section, the S\&I operator is built from the local Hamiltonian for the site on which the optimization is performed. This means that the operator (i.e., the left and right boundaries) changes at each iteration step, and therefore a different value of $\omega$ can be employed in each microiteration step. Here, we propose to set $\omega = \omega_\text{pr} - \omega_\text{shift}$, where $\omega_\text{pr}$ is the energy of the MPS at the previous microiteration step, and $\omega_\text{shift}$ becomes a parameter of the algorithm. 
$\omega_\text{shift}$ is introduced to avoid instabilities in the definition of S\&I operator, which diverges when $\omega$ is equal to one of the eigenvalues and to take into account that the energy will, in general, decrease after a microiteration step.
The dynamical update of $\omega$ is not possible for DMRG[f], in which the shift parameter is included in the definition of the MPO. However, as will be discussed in the application section, for DMRG[f] the choice of $\omega$ has a little impact on the convergence of DMRG compared to the other variants.

All algorithms introduced require the calculation of the overlap between vibrational MPSs. 
The overlap between two MPSs $\ket{\Psi^{(k)}}$ and $\ket{\Phi^{(h)}}$ for states $k$ and $h$, respectively, can be calculated as\cite{scho11,Keller2015_MPS_MPO}

\begin{align}
  \OvI{\Psi^{(k)}}{\Phi^{(h)}} = \sum_{\sigma_L}^{\bm{N}_\text{max}} \bm{M}^{\sigma_L (k) \dagger} 
         \cdots \nonumber \qquad \qquad \qquad \\ 
  \left( \sum_{\sigma_1}^{\bm{N}_\text{max}} \bm{M}^{\sigma_1 (k) \dagger} \bm{N}^{\sigma_1 (h) } \right) 
                          \cdots \bm{N}^{\sigma_L (h) } \, , 
  \label{eq:VMDRG_overlap}
\end{align}
where the $\bm{M}^{\sigma_i(k)}$ matrices are associated with $\ket{\Psi^{(k)}}$, whereas the $\bm{N}^{\sigma_i (h)}$ define $\ket{\Phi^{(h)}}$.

At each iteration step of the Jacobi-Davidson (or Davidson) algorithm, $N_\text{states}$ lowest-energy roots are calculated (corresponding to $N_\text{states}$ different MPSs), and their overlap with $\ket{\Phi_\text{test}}$ is calculated. 
The MPS with the largest overlap is tracked, and used in the subspace expansion step (Eq.~(\ref{eq:JD_correction_modified_2})).

As already discussed,\cite{Keller2015_MPS_MPO} there is no need for calculating the overlap from Eq.~(\ref{eq:VMDRG_overlap}) in each microiteration step, because only one $\bm{M}^{\sigma_i}$ matrix is optimized at a time. It is convenient to introduce the partial overlap matrices,
\begin{align}
    \bm{C}^l =& \sum_{\sigma_l}^{N_\text{max}} \bm{M}^{\sigma_l \dagger} \cdots
        \left( \sum_{\sigma_1}^{N_\text{max}} \bm{M}^{\sigma_1 \dagger} \bm{N}^{\sigma_1} \right)
               \ldots \bm{N}^{\sigma_l} 
  \label{eq:Partial_overlap_1} \\
    \bm{D}^l =& \sum_{\sigma_{l+1}}^{N_\text{max}} \bm{N}^{\sigma_{l+1}} \cdots
        \left( \sum_{\sigma_L}^{N_\text{max}} \bm{N}^{\sigma_L} \bm{M}^{\sigma_L \dagger} \right) 
               \ldots \bm{M}^{\sigma_{l+1} \dagger} \, , \\
  \label{eq:Partial_overlap}
\end{align}
from which it is easy to show\cite{Keller2015_MPS_MPO}
\begin{align}
  \OvI{\Psi}{\Phi} = \text{tr} \left( \bm{C}^l \bm{D}^l \right) \, .
\end{align}
The $\bm{C}$ and $\bm{D}$ vectors of matrices are stored during the optimization. 

To conclude, we emphasize that the combination of the DMRG[MaxO] algorithms with the Davidson and Jacobi-Davidson diagonalization is particularly appealing. 
In fact, as already discussed above, the S\&I algorithm converges to the eigenvalue that is the closest to the $\omega$ parameter. 
However, when targeting states localized in regions with a high density of states, the interval of values of $\omega$ in which the optimization converges to the targeted states might be very small and it might be difficult to set $\omega$ appropriately. 
The choice of $\omega$ is less critical in DMRG[MaxO], because several eigenstates are approximately calculated at each iteration step, and the MaxO criterion allows us to consistently optimize the state of interest.

\subsection{Multi-state DMRG}

As already noted in Ref.~\citenum{Keller2015_MPS_MPO} for the MPO-MPS formulation of DMRG, it is possible to obtain excited states also by optimizing the MPS in the space orthogonal to all the lower-energy states.
For example, the orthogonality of the first excited state $\ket{\Psi^{(1)}}$ with respect to the ground state $\ket{\Psi^{(0)}}$ when optimizing the MPS on the $l$-th site can be easily expressed by introducing a matrix $\bm{V}^{l(0)}$ defined as
\begin{equation}
  \bm{V}^{l(0)} = \sum_{\sigma_l} \bm{C}^{l-1} \bm{N}^{\sigma_l(0)} \bm{D}^l \, ,
  \label{eq:OrthoVector}
\end{equation}
where the definition of $\bm{C}^{l-1}$ and $\bm{D}^l$ is as in Eqs.~(\ref{eq:Partial_overlap_1}) and (\ref{eq:Partial_overlap}). 
One can show that, by keeping the $\bm{M}^{\sigma_l(1)}$ matrix orthogonal to $\bm{V}^{l(0)}$ during the $l$-th microiteration step that the orthogonality constraint between the MPSs is fulfilled.\cite{McCulloch2007_FromMPTtoDMRG,Keller2015_MPS_MPO}

In the present work, we apply a modified version of our initial algorithm\cite{Keller2015_MPS_MPO,Baiardi2017_vDMRG} for calculating excited states, for which it was necessary to fully optimize all lower $n-1$ states to optimize the $n$-th excited state.
As already noted above, this task can be challenging in regions with a high density of states, where root-flipping is commonly observed. 
Orthogonality can, however, be ensured by calculating the $\bm{V}^l$ vectors for the first $n$ states \textit{on the fly} in each microiteration step.
We refer to this modified approach as orthogonal multi-state DMRG (DMRG[oMS]).

\subsection{Stochastic sampling of the occupation number vector space}

An unavoidable drawback of DMRG compared to CI approaches is that the wave function $\ket{\Psi^{(k)}}$ is expressed as an MPS, and hence, all CI coefficients cannot be known for the algorithm to be efficient. 
For this reason, it is not easily possible to determine the configurations with the largest coefficients in the CI expansion. For vibrational wave function, for example, this means that it is not possible to characterize the MPS as a fundamental, an overtone, or a combination band.
The theoretical framework outlined in the previous subsections allows us to calculate the overlap between a single ONV and an MPS, and can be applied to reconstruct the CI wave function by calculating the overlap with all possible ONVs. 
Such an algorithm has already been implemented in Ref.~\citenum{Moritz2007_CIReconstruction}, but is limited to very small systems, due to the exponential increase of the variational space with the number of DMRG sites. 
To handle also larger systems, a more efficient way to sample the variational space is required.
Here, the sampling reconstruction complete active space (SR-CAS) algorithm, developed for the electronic structure problem in Ref.~\citenum{Boguslawski2011_SRCAS} offers a remedy. Other algorithms for sampling the ONV space have been proposed, based either on Monte Carlo-based techniques\cite{Vidal2012_PerfectSampling} or on genetic algorithms.\cite{Ma2017_CASCIReconstructionEntanglment,Ma2018_ExternallyContractedMRCI-DMRG} All these algorithms can be extended to vDMRG as well.

In the SR-CAS algorithm, the variational space is sampled through a Metropolis-Hastings Markov chain,\cite{Chib1995_MetropolisHastings} where the probability density $\rho_{\sigma_1,...,\sigma_L}$ is given by the squared value of the CI coefficient $C_{\sigma_1,...,\sigma_L}$,
\begin{equation}
  \rho_{\sigma_1,...,\sigma_L} = \left| C_{\sigma_1,...,\sigma_L} \right|^2 \, .
  \label{eq:Metropolis_distribution}
\end{equation}
$\rho_{\sigma_1,...,\sigma_L}$ can be interpreted as a probability distribution function because $\sum_{\sigma_1} ... \sum_{\sigma_L} \rho_{\sigma_1,...,\sigma_L} = 1$. 
The Metropolis-Hastings algorithm is designed to sample regions of the CI space with higher probability density more often and, hence, avoids the repeated calculation of almost negligible overlaps.
The convergence of the algorithm can be easily assessed through the completeness measure $\text{COM}$ defined as
\begin{equation}
  \text{COM} = 1 - \sum_{i} \left| C_i \right|^2 ,
  \label{eq:Completeness_measure}
\end{equation} 
where the sum over $i$ includes only the stored determinants. Without going into the details of the algorithm, which are reported in the SI, we here generalize SR-CAS\cite{Boguslawski2011_SRCAS} to vibrational Hamiltonians. The only step of the original SR-CAS that must be modified is in the generation of the new ONVs during the random sampling. In electronic structure theory only four possible occupations (unoccupied, spin up, spin down, doubly occupied) are possible for spatial orbitals, whereas for molecular vibrations the occupation number is limited only by the $N_\text{max}$ parameter.

As proposed by Carrington and co-workers in the context of CP factorization,\cite{Carrington2018_CP-RRPM} the ONV with the largest configuration can be identified also by compressing the rank of the optimized wave function (i.e., $m$ in DMRG) up to rank 1. However, unlike SR-CAS, this procedure enables one to obtain only the predominant ONV, and not to reconstruct the full expansion of the MPS in terms of the CI basis up to a given accuracy.

\section{Computational details}
\label{sec:details}

We apply the theory presented in the previous section to the calculation of vibrational energies of molecular systems. We take as reference the Watson-type Hamiltonian $\mathcal{H}_\text{vib}$ already employed in our previous work.\cite{Baiardi2017_vDMRG} We employ Cartesian normal modes as the reference coordinate system, and expand the potential energy operator as a Taylor series around some reference structure including up to sixth-order terms. Even our original vDMRG implementation includes also first-order Coriolis couplings in the kinetic energy operator, in the present work we neglect ro-vibrational coupling terms. The MPO form of $\mathcal{H}_\text{vib}$ can be built starting from its canonical second-quantization form, which can be found, for example, in Ref.\citenum{Baiardi2017_vDMRG}.

We implemented all algorithms presented in the previous sections to target vibrationally excited states in our \textsc{QCMaquis-V} program,\cite{Baiardi2017_vDMRG} which was derived from the \textsc{QCMaquis} program\cite{Keller2015_MPS_MPO,knec16} written for electronic structure calculations.

The anharmonic force fields applied in the vDMRG calculations were taken from the literature for C$_2$H$_4$\cite{Delahaye2014_EthylenePES} or were calculated with the \textsc{Gaussian} program\cite{g16.a03} in the case of the sarcosine-glycine dipeptide, {SarGly$^+$}.
Detailed information on the electronic structure methods applied for the generation of the force-field is given in the respective sections.

We emphasize that the variational optimization of the ground-state MPS provides the anharmonic zero-point vibrational energy (ZPVE). 
All the algorithms presented in the previous section deliver the absolute energy of a vibrational state. 
Transition energies $h\nu_k$ are then calculated as
\begin{align}
  h\nu_k = E_{k}-\text{ZPVE} \, .
  \label{trans_energy}
\end{align}
If not otherwise specified, in vDMRG[S\&I] calculations the $\omega_\text{shift}$ parameter was set to 10~cm$^{-1}$.

For SR-CAS calculations, the $\eta$ threshold for the completeness COM was set to $10^{-3}$ if not otherwise specified.
The Davidson and Jacobi-Davidson convergence threshold was set to 0.1~cm$^{-1}$ in all cases, and a maximum of 40 iterations of the subspace iteration algorithm was employed.
The threshold for assessing the convergence of the correction equation was set to $\left| \bm{r}_n \right| / 10$ in all cases, where $\bm{r}_n$ is the error vector calculated in the $n$-th iteration of the Jacobi-Davidson diagonalization.
Such a threshold does not lead to the exact solution of the correction equation but, as discussed in Ref.~\citenum{Sleijpen2000_JacobiDavidson}, an exact solution is not mandatory for this equation.

In all cases, the optimization of MPSs was carried out with a single-site optimizer, hence optimizing one $\bm{M}^{(k)}$ tensor per microiteration. As already discussed in the literature,\cite{Schollwoeck2005,scho11} single site optimization algorithms can lead to slow convergence, or even to convergence to local minima. We follow the approach described in Ref.~\citenum{McCulloch2015_Mixing}, where the reduced density matrix is perturbed by a noise term before the truncation to speed-up the convergence rate of the algorithm. If not otherwise specified, the perturbation parameter $\alpha$ was set to 10$^{-8}$ for the first 10 sweeps and then set to zero for the remaining macroiterations. Our vDMRG implementation supports also a two-site optimization algorithm,\cite{scho11} which is an alternative route to speed-up the convergence of DMRG. However, in this work, only the single-site optimizer will be employed since, as shown in Fig.~S6 for the $\nu_{11}$ vibration of ethylene, the convergence rate of the two methods is equivalent.

\section{Applications}

\begin{figure}
	\centering
	\includegraphics[width=.4\textwidth]{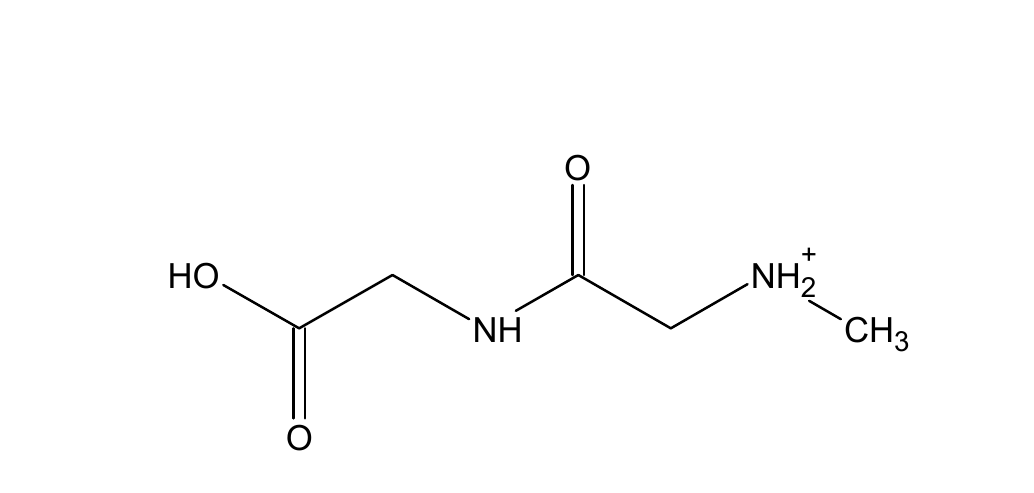}
	\caption{Lewis structure of SarGly$^+$.}
	\label{fig:Dipeptide}
\end{figure}

The energy-specific variants of vDMRG presented in the previous section are applied to ethylene and to the dipeptide SarGly$^+$, whose structure is reported in Fig.~\ref{fig:Dipeptide}.
These systems have already been studied with the standard variant of vDMRG in our previous work\cite{Baiardi2017_vDMRG} and hence provide reference data, against which  the various algorithms introduced above can be compared.
However, in the present study we also show how the energy-specific variant of vDMRG allows to target highly-excited states, with energies above 3000~cm$^{-1}$, with computational cost comparable to the one of the standard variant of vDMRG for ground states. 
Targeting these excited states with standard vDMRG is impossible, because of the steep increase of computational cost due to the huge number of lower energy excited states.

\subsection{Ethylene}

Ethylene is our first example to highlight the capabilities of the energy-specific variants of vDMRG presented in the Section~\ref{sec:theory}.
From our previous work, we adopt a quartic force fields generated from a recently published\cite{Delahaye2014_EthylenePES} highly accurate potential energy surface in Cartesian normal coordinates.

\subsubsection{Root-homing in vDMRG}

\begin{figure}
	\centering
	\includegraphics[width=.5\textwidth]{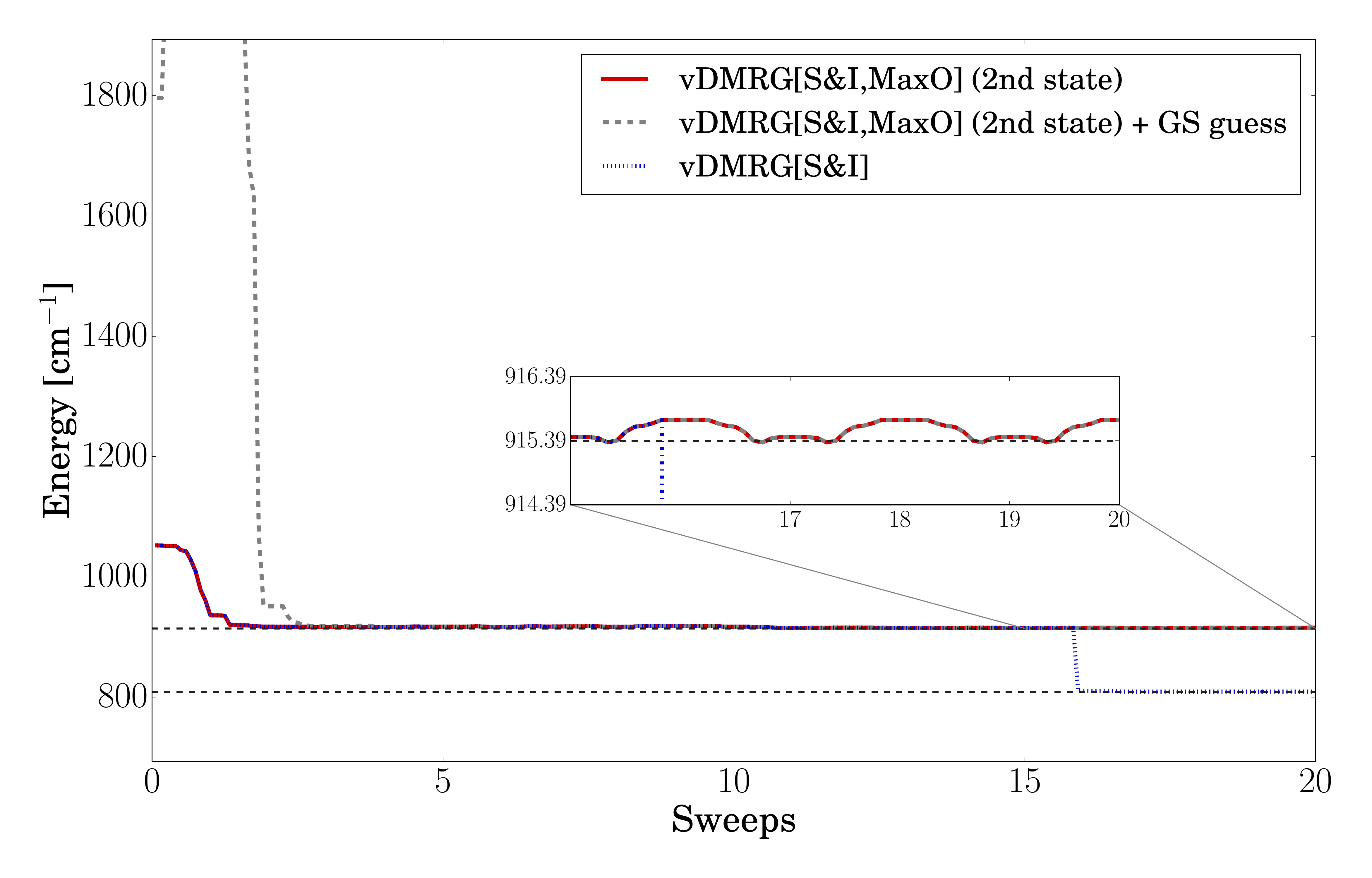}
	\caption{vDMRG transition energies as a function of the number of sweeps in which each microiteration step applied the S\&I Jacobi-Davidson algorithm and an energy shift $\omega$ of 11800 cm$^{-1}$. Staring from the ground (dashed gray line) and second excited (solid red line) vibrational harmonic levels as a guess and the ONV corresponding to the harmonic wave function of the second excited state as test MPS, we set $m$=20 and $N_\text{max}$=6. The converged energy of the first and second vibrationally excited states (11817.28~cm$^{-1}$ and 11921.38~cm$^{-1}$ respectively) are indicated by a horizontal dashed black line. Three eigenstates were kept during the tracking with the MaxO algorithm.}
  \label{fig:Convergence_MOM}
\end{figure}

The Jacobi-Davidson diagonalization requires on average about half as many iteration steps as the Davidson eigensolver (as demonstrated in detail in the Supplementary Material) and will therefore be applied in all following calculations.
In these calculations, the converged vDMRG energies were known \textit{a priori} and therefore allowed us to set the shift $\omega$ for the target eigenstate to be the lowest-energy eigenstate of the S\&I operator $\Omega_\omega$. 
Obviously, converged vDMRG energies are, in general, not known, and only an estimate of the vibrational energies (for example their harmonic value) is available. 
In such cases, vDMRG[MaxO] is particularly appealing, because it allows one to track a selected state during the optimization based on its overlap with a trial wave function.
To highlight the strengths of vDMRG[MaxO], we set $\omega$ to 11800 cm$^{-1}$ and target the second vibrational excited state.
Standard vDMRG[S\&I] will converge to the first excited state, and optimization of the second excited state would again require a subsequent constrained optimization.
However, as shown in Fig.~\ref{fig:Convergence_MOM}, if the eigenstate with the maximum overlap with the harmonic wave function of the second excited state is tracked, the optimization converges to the second vibrational level without the need of any further constrained optimization. 
Even when the optimization is started from the vibrational ground state as a guess (gray dashed line in Fig.~\ref{fig:Convergence_MOM}), the algorithm converges to the correct asymptotic value. 
By contrast, without root-homing, the optimization oscillates for several sweeps around the energy of the second excited state, but then it converges, as expected, to the first vibrational excited state. 

\begin{table*}[ht!]
	\begin{center}	
		\begin{tabular}{cc|cccccccc|c}
			\hline\hline
			State & Assignment & $m = 10$ & $m = 20$ & $m = 30$ & $m = 40$ & $m = 50$ & $m = 60$ & $m = 100$ & Ref.\cite{Baiardi2017_vDMRG} &  $\omega$   \\
			\hline
           ZPVE  &             & 
              11008.65 & 11006.61 & 11006.32 & 11006.22 & 11006.18 & 11006.16 & 11006.13  & & \multirow{4}{*}{$\omega$ = 11000 cm$^{-1}$} \\
			1    &  $\nu_{10}$          &   811.75 &   809.45 &   808.91 &   808.80 &   808.65 &   808.65 &   808.53  & 809.03 & \\
			2    &  $\nu_9$             &   917.41 &   915.72 &   915.29 &   915.09 &   914.75 &   914.94 &   914.99  & 915.29 & \\
			3    &  $\nu_8$             &   930.45 &   928.79 &   928.13 &   928.11 &   928.08 &   928.01 &   927.91  & 928.31 & \\
			\hline 
			4    &  $\nu_4$             &  1009.45 &  1007.60 &  1007.12 &  1006.95 &  1006.93 &  1006.80 &  1006.76  & 1007.03 &  \multirow{4}{*}{$\omega$ = 12000 cm$^{-1}$} \\
			5    &  $\nu_6$             &  1218.52 &  1217.41 &  1217.14 &  1217.02 &  1216.96 &  1216.91 &  1216.87  & 1217.17 & \\  
			6    &  $\nu_3$             &  1340.31 &  1339.23 &  1338.78 &  1338.61 &  1338.52 &  1338.48 &  1338.41  & 1338.87 & \\  
			7    &  $\nu_{12}$          &  1433.46 &  1430.96 &  1430.42 &  1430.16 &  1429.98 &  1429.96 &  1429.88  & 1430.47 & \\  
			\hline
			8    &  $\nu_2$             &  1618.62 &  1609.75 &  1607.49 &  1606.59 &  1606.23 &  1606.01 &  1605.37  & 1622.11 &  \multirow{4}{*}{$\omega$ = 12500 cm$^{-1}$} \\
			9    & 2$\nu_{10}$          &  1634.82 &  1634.97 &  1632.47 &  1632.23 &  1631.94 &  1631.16 &  1630.19  & 1625.56 &  \\   
			10   & $\nu_8 + \nu_{10}$   &  1725.70 &  1721.00 &  1727.13 &  1720.10 &  1719.24 &  1718.56 &  1717.76  & 1722.77 & \\  
			11   & $\nu_7 + \nu_{10}$   &  1741.19 &  1736.52 &  1734.67 &  1734.32 &  1735.52 &  1734.86 &  1733.59  & 1729.53 & \\ 
			\hline\hline
		\end{tabular}
		\caption{Vibrational energies (in cm$^{-1}$) of the twelve lowest vibrational levels (numbered in energetic order, the assignment and their harmonic frequencies are reported in Table~S1 of the Supplementary Material) of ethylene calculated from a quartic force-field in Cartesian normal coordinates with different numbers of renormalized block states $m$. In all cases, $N_\text{max}$ was set to 6. The S\&I variant of vDMRG was employed in all cases, with three different values for $\omega$ (11000, 12000, and 12500 cm$^{-1}$) as indicated in the right column. For each energy shift value, the first four lowest states were calculated. In all cases, vDMRG[MaxO] was employed with a test MPS, in which only the ONV in the second column does not vanish, to calculate the overlaps.}
       	\label{tab:C2H4_Multistate}
	\end{center}
\end{table*}

In Table~\ref{tab:C2H4_Multistate}, the energy of the twelve lowest states of ethylene obtained with the state-specific variant of vDMRG with the S\&I algorithm are reported as a function of the number of renormalized block states $m$. 
Following the ideas reported above, the energies were obtained with three separate S\&I calculations, corresponding to three different values of the shift parameter $\omega$ (11000, 12000, 12500 cm$^{-1}$). 
In all cases, the state-specific variant of vDMRG was employed to calculate the first four lowest-energy roots. 
Calculations were performed with different values of $m$, ranging from 10 to 100. 
First of all, we note that, in agreement to what we found in our previous work,\cite{Baiardi2017_vDMRG} $m=20$ is sufficient to reach convergence within 1 cm$^{-1}$ for all the states. 
Furthermore, all data are in good agreement with the results reported in Ref.~\citenum{Baiardi2017_vDMRG} (with variation below 1 cm$^{-1}$) that were calculated with the state-specific vDMRG variant, without the S\&I algorithm. 
Only for the 8th excited state, the deviation amounts to 17~cm$^{-1}$. As reported in Table~\ref{tab:C2H4_Multistate}, states 8 and 9 are close in energy and the corresponding anharmonic wave functions strongly deviates from the harmonic reference. Under these conditions, \textit{i.e.} in presence of closely-lying strongly coupled states, the state-specific variant of vDMRG employed in our reference paper\cite{Baiardi2017_vDMRG} is prone to get stuck in local minima. For this reason, the difference can be ascribed to an incomplete convergence of the results reported in Ref.~\citenum{Baiardi2017_vDMRG}.

\begin{figure}[h!]
	\centering
	\includegraphics[width=.5\textwidth]{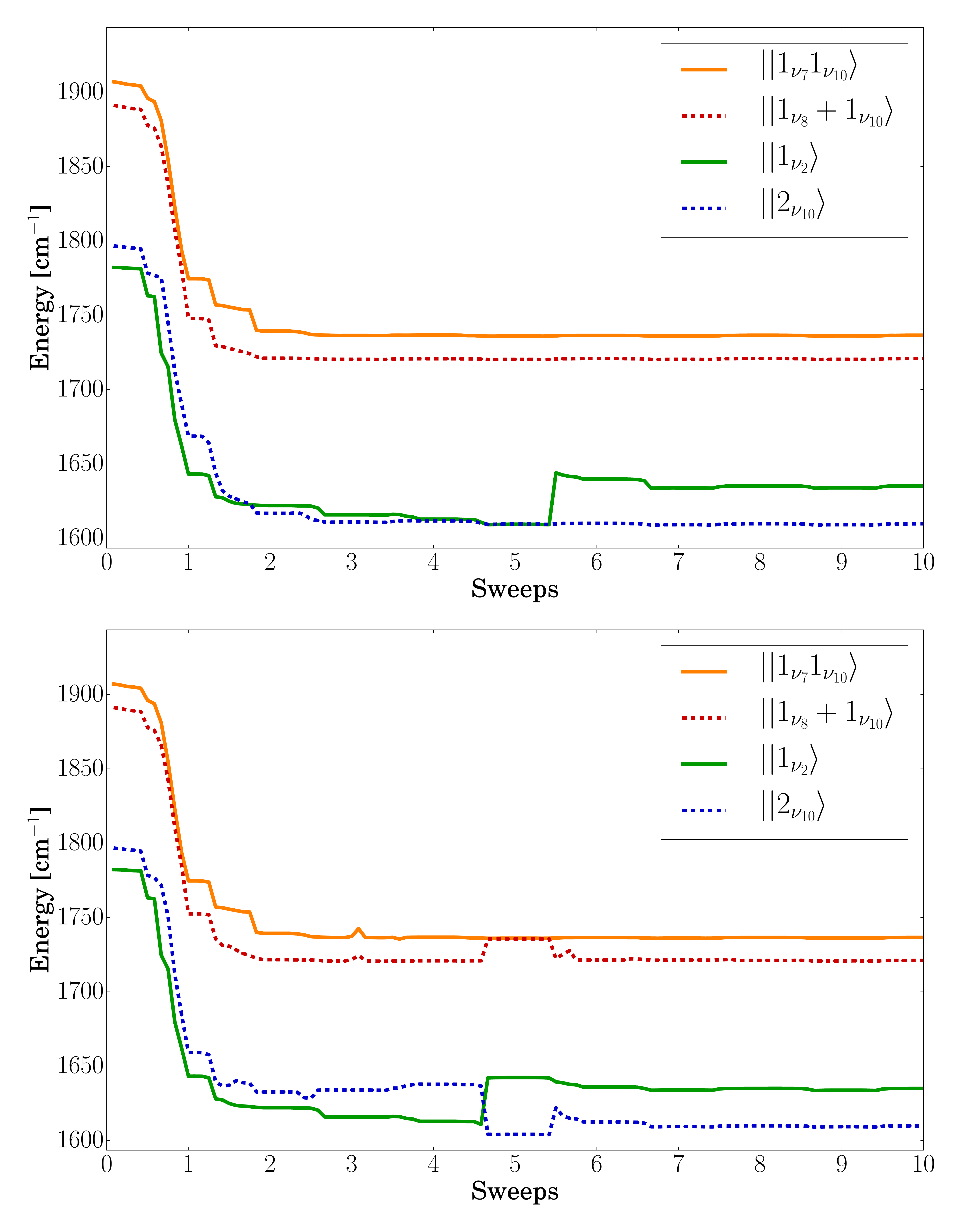}
	\caption{Convergence of states 8 to 11 (numbered in increasing energetic order) of C$_2$H$_4$ (see Table~\ref{tab:C2H4_Multistate}) as a function of the number of sweeps. Vibrational transition energies are reported. The vDMRG parameters are $m=20$ and $N_\text{max}=6$. The vDMRG[MaxO] variant was chosen with $\omega$ = 12500 cm$^{-1}$. Results reported in the upper panel are obtained with a separate calculation for each state without any orthogonalization step. The ones reported in the lower panel are obtained with vDMRG[oMS]. Vibrational states are labeled according to the test MPS for which the overlap was calculated.}
	\label{fig:Conv_8and9_C2H4_MO}
\end{figure}

\subsubsection{Anharmonic coupling, root flipping, and multi-state vDMRG}
Strong anharmonic coupling for two vibrational states close in energy are challenging for the root-homing algorithm.
This is due to the coupling that introduces a strong mixing of the wave functions that serve as an initial guess and contain only one non-vanishing ONV.
The shift parameter $\omega$ does not offer a solution in those cases where the states are very close in energy (e.g., for resonant vibrational states).
We illustrate this fact for the $\nu_2$ and $2\nu_{10}$ states of ethylene in the upper panel of Fig.~\ref{fig:Conv_8and9_C2H4_MO}, where both states appear to converge to the same energy (until the 5-th sweep) because the overlap criterion alone does not differentiate them.
We show results of the same calculation but with the oMS-vDMRG algorithm in the lower panel of Fig.~\ref{fig:Conv_8and9_C2H4_MO}.
Here, the orthogonalization enforces each root to correspond to a different state instead of convergence to the same energy and state.
Root flipping can still occur but only in pairs (cf. the 5-th sweep) and is hence better described as a root exchange.
Concerning the pair of states $\nu_8+\nu_{10}$ and $\nu_7+\nu_{10}$, the overlap criterion is sufficient because there is no strong mixing of the corresponding harmonic guess wave functions.

\subsubsection{Sampling reconstruction of the determinant space}
\begin{figure}
	\centering
	\includegraphics[width=.5\textwidth]{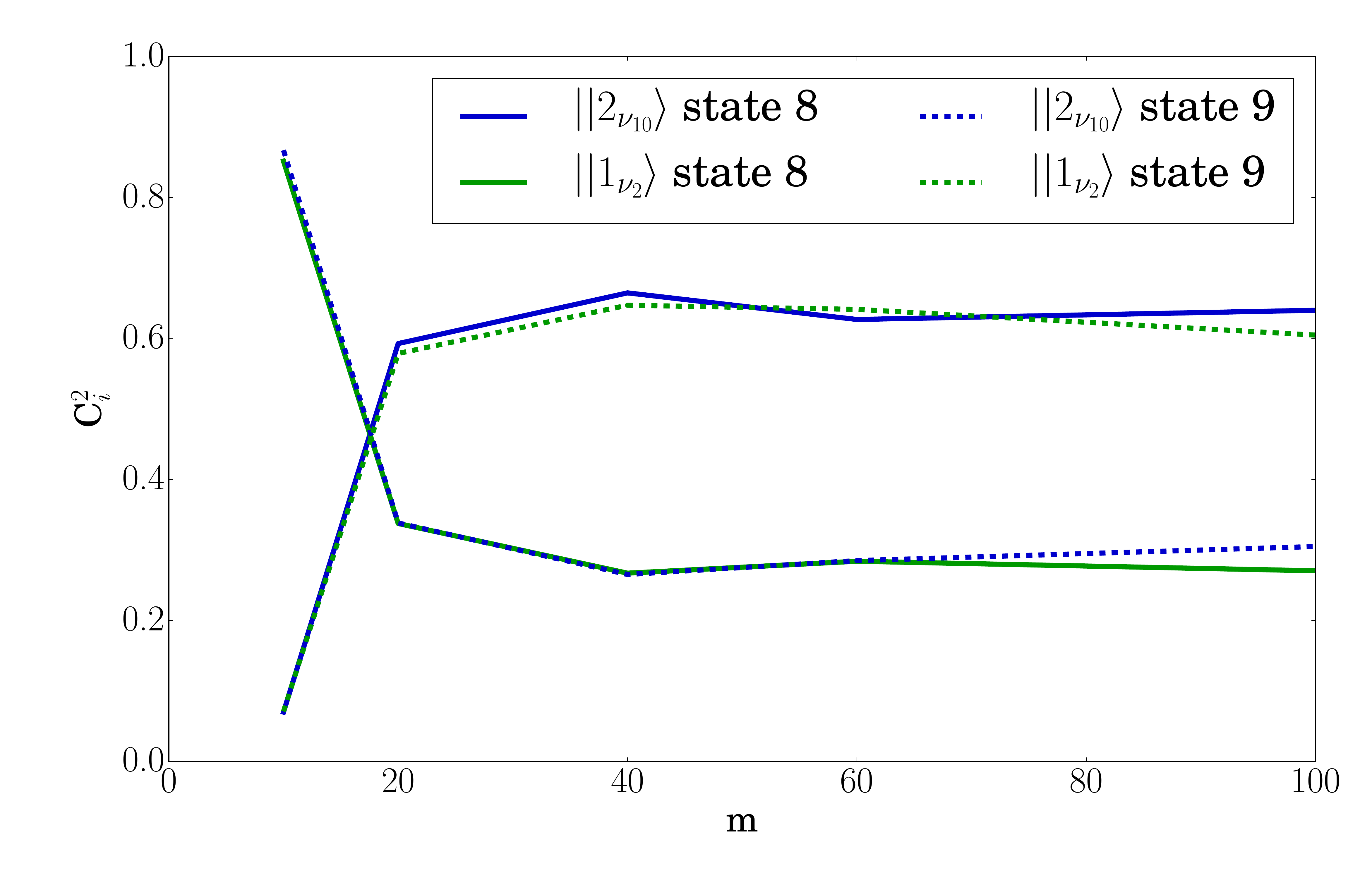}
	\caption{Graphical representation of the results of the SR-CAS algorithm applied to the MPSs resulting from the optimization starting from the $\nu_2$ fundamental (solid green line, labeled as state 8) and the $2\nu_{10}$ overtone (dashed blue line, labeled as state 9) of C$_2$H$_4$ as test MPS. The coefficients of the $\nu_2$ fundamental (green lines) and of the $2\nu_{10}$ overtone (red lines) are reported. All calculations were carried out with a quartic force-field, with $N_\text{max}=6$. SR-CAS calculations were carried out with $\eta=10^{-3}$.}
	\label{fig:C2H4_SR-CAS}
\end{figure}

From Fig.~\ref{fig:Conv_8and9_C2H4_MO}, it is clear that there is a significant difference between the convergence of the pair of lower-energy states ($||1_{\nu_2}\rangle$ and $||2_{\nu_{10}}\rangle$, named after the ONV with the largest coefficient) and the pair of higher-energy states ($||1_{\nu_8}1_{\nu_{10}}\rangle$ and $||1_{\nu_7}1_{\nu_{10}}\rangle$).
For the latter pair of states, the CI coefficient of the guess in the final wave function is predominant, being larger than 0.975 in both cases.
For the two lower-energy states, the CI coefficient of the starting guess in the final wave functions is much smaller, lower than 0.8 in both cases. 
This means that, because of strong anharmonic effects, several basis states have a large coefficient in the CI expansion. 
We analyzed the configurations with largest weights with the SR-CAS algorithm. The results of the SR-CAS algorithm are shown in Fig.~\ref{fig:C2H4_SR-CAS}, where the coefficients of the ONV associated with the fundamental $\nu_2$ and the overtone $2\nu_{10}$ are reported. 
First of all, the results of the SR-CAS algorithm indicate that both states have a large ($>$0.1) weight for both ONVs $||1_{\nu_2}\rangle$ and $||2_{\nu_{10}} \rangle$. 
Furthermore, the variation of the coefficients of the two CI states with the number of renormalized block states $m$ is symmetric for the two states, i.e. the variation of the coefficient of $|| 1_{\nu_2} \rangle$ for state 8 is equivalent to the one of $|| 2_{\nu_{10}} \rangle$ for state 9 and vice versa, with the states being numbered according to their energetic order. 
This trend indicates that only those two harmonic states have a non-negligible contribution to the converged CI expansion. 
In fact, in this case, the vibrational wave functions $\ket{\Psi^{(8)}}$ and $\ket{\Psi^{(9)}}$ can be well approximated as:

\begin{equation}
	\begin{aligned}
		\ket{\Psi^{(8)}} \approx C_{\nu_2}^{(8)} ||1_{\nu_2}\rangle &+ C_{2\nu_{10}}^{(8)} ||2_{\nu_{10}}\rangle \\
		\ket{\Psi^{(9)}} \approx C_{\nu_2}^{(9)} ||1_{\nu_2}\rangle &+ C_{2\nu_{10}}^{(9)} ||2_{\nu_{10}}\rangle \\
	\end{aligned}
	\label{eq:RabiModel4DMRG}
\end{equation}
and, from the orthogonality constraint $\OvI{\Psi^{(8)}}{\Psi^{(9)}} = 0$, it follows $\left|C_{\nu_2}^{(8)}\right| = \left|C_{2\nu_{10}}^{(9)}\right|$ and $\left|C_{\nu_2}^{(9)}\right| = \left|C_{2\nu_{10}}^{(8)}\right|$. 
We also note that the convergence of the energy is faster than that of these coefficients. 
In fact, although with $m=20$ a near complete convergence of the energy is reached, as shown in Table~\ref{tab:C2H4_Multistate}, significant variations in the wave function composition are still observed.

\subsubsection{Shift and inverse algorithm for high-energy states}

Up this point, we have employed S\&I algorithm to target vibrational modes, that were already known from our previous work.\cite{Baiardi2017_vDMRG} To highlight the robustness of vDMRG[S\&I], the algorithm is applied to target the four C-H stretching modes. Based on the theoretical results given in Ref.~\citenum{Delahaye2014_EthylenePES}, the first C-H stretching mode is mode No.~63 (for states ordered by increasing energy). Hence, its calculation with standard state-specific algorithms would require the optimization of all the 62 lower vibrational states. 
This task is extremely challenging, not only because of the high computational cost, but also because the region between 2000-3000 cm$^{-1}$ shows a high density of states. 
Furthermore, according to the results reported in Ref.~\citenum{Delahaye2014_EthylenePES}, the first C-H stretching mode features an anharmonic frequency of 2976 cm$^{-1}$, and the first lower- and higher-energy states are found at frequencies of 2972 and 2991 cm$^{-1}$, respectively. 
Hence, in standard S\&I approaches, by choosing $\w$ between 2972 and 2991 cm$^{-1}$ facilitates convergence to the second stretching mode. However, as will be discussed in the following, the MaxO variant of vDMRG makes the choice of $\w$ less critical in this case.
 
The first C-H stretching mode of ethylene belonging to the $B_{1u}$ point group (referred to as $\nu_{11}$ in the following, according to the notation in Ref.~\citenum{Delahaye2014_EthylenePES}), was calculated to assess the reliability of the different vDMRG[MaxO] variants.
To limit the computational cost, the quartic force-field in Cartesian normal coordinates taken from Refs.~\citenum{Delahaye2014_EthylenePES} and \citenum{Crittenden2015_PyPES} was employed. 
Calculations with the more accurate, sixth-order force-field are presented in the next section. 
As the $\nu_{11}$ mode lies in an energy range with a high density of states, the S\&I variant of vDMRG is incapable of targeting this mode (as shown in Fig.\ 2 of the Supplementary Material), because only if $\omega$ is chosen in a very narrow energy range, convergence to the correct state will be obtained. 
The absolute energy of this mode reported in Ref.~\citenum{Delahaye2014_EthylenePES} is 14000.29 cm$^{-1}$. 
Since $\omega$ should be lower than the energy of the targeted states, calculations with vDMRG[S\&I] were performed with $\omega$ = 13900, 13950 and 13980 cm$^{-1}$ (results are reported in Fig.~S2 of the Supplementary Material).
In all cases, the optimization algorithm did not converge after 20 sweeps. 
This demonstrates that, in regions with a high density of states, the S\&I variant is not sufficient to ensure convergence to either the target vibrational state or to any state.

\begin{figure}[h!]
	\centering
	\includegraphics[width=.5\textwidth]{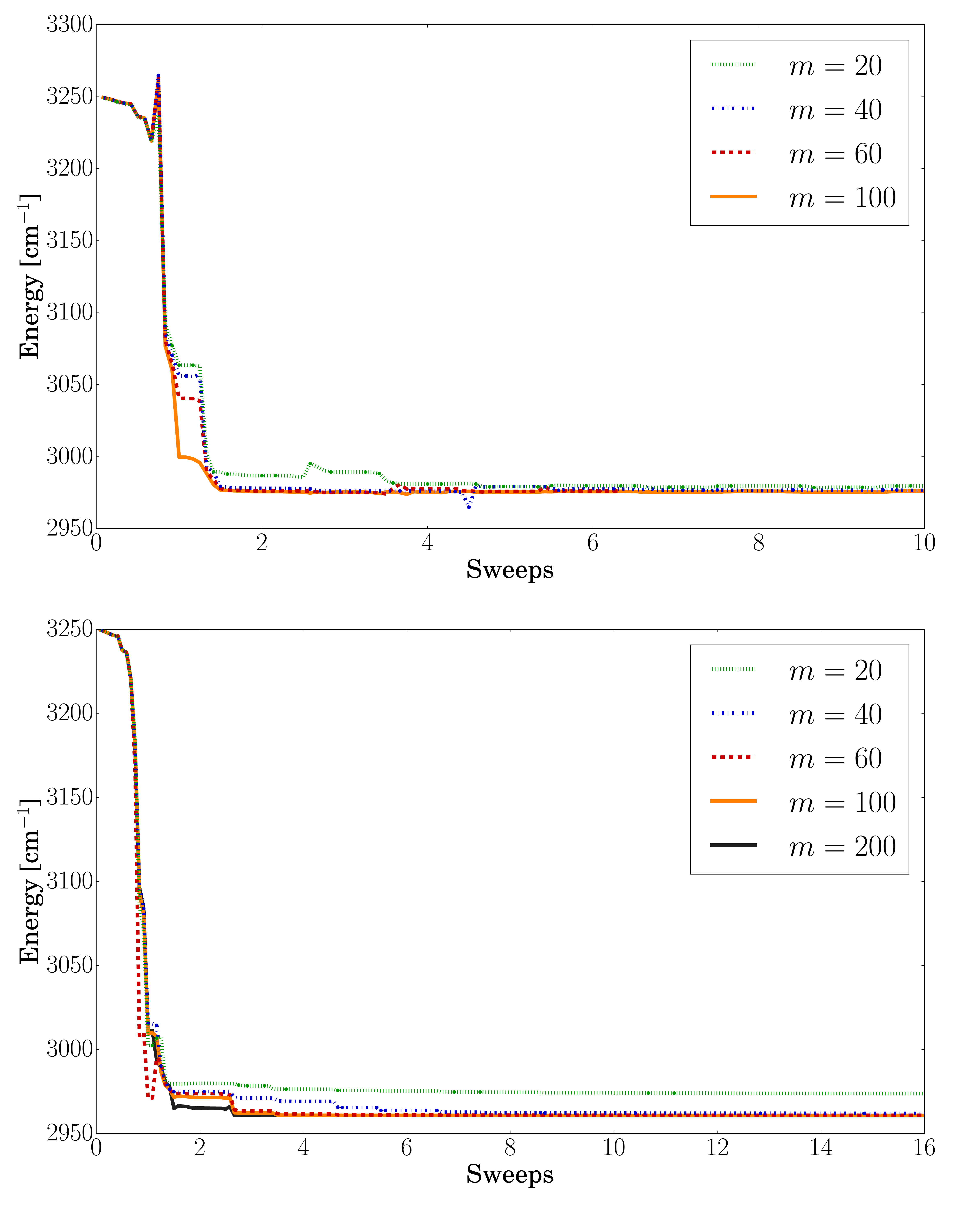}
	\caption{vDMRG[S\&I,MaxO] (upper panel) and vDMRG[f,MaxO] (lower panel) energies as a function of the number of sweeps of the $\nu_{11}$ mode of ethylene for different values of the number renormalized block states $m$. The $\omega$ parameter was updated with $\omega_\text{shift}$ = 10~cm$^{-1}$.}
	\label{fig:CH1_convergence}
\end{figure}

To increase the accuracy of vDMRG[S\&I], its MaxO formulation has been employed and applied to the $\nu_{11}$ mode. 
The plot of the vibrational energy as a function of the number of sweeps, reported in the upper panel of Fig.~\ref{fig:CH1_convergence}, shows that combining the Jacobi-Davidson iterative solver with root-homing allows to converge the DMRG optimization. For all values of $m$, convergence is reached with a relatively small number of renormalized states ($m = 60$) and variations below 1 cm$^{-1}$ are detected with $m=100$. This indicates that the rate of convergence with respect to $m$ is slower than for the lower energy states. 
However, only a slight increase in the value of $m$ leads to a  complete convergence of a highly excited state, such as the one involved in the $\nu_{11}$ transition.
This also supports the analysis reported in our previous work,\cite{Baiardi2017_vDMRG} where large variations in the ZPVE between $m$=20 and 40 renormalized states was assumed to be associated to an incomplete convergence of the energy of the C-H stretching modes. 

The S\&I transformation was applied only to an approximate representation of the Hamiltonian. A more robust alternative is vDMRG[f], where the spectral transformation is applied to the full Hamiltonian. We study vDMRG[f] for the optimization of the $\nu_{11}$ vibrational state, employing the same parameters reported for vDMRG[S\&I]. The results are reported in the lower panel of Fig.~\ref{fig:CH1_convergence}. As expected, also in this case convergence is reached within 6 sweeps, and changes in the energies below 1~cm$^{-1}$ are observed for values of $m$ higher than 40. It is worth noting that, for higher values of $m$ (100 and 200), the optimization is more efficient than with lower values of $m$, and convergence is obtained within 3-4 sweeps.

\begin{figure}[h!]
	\centering
	\includegraphics[width=.5\textwidth]{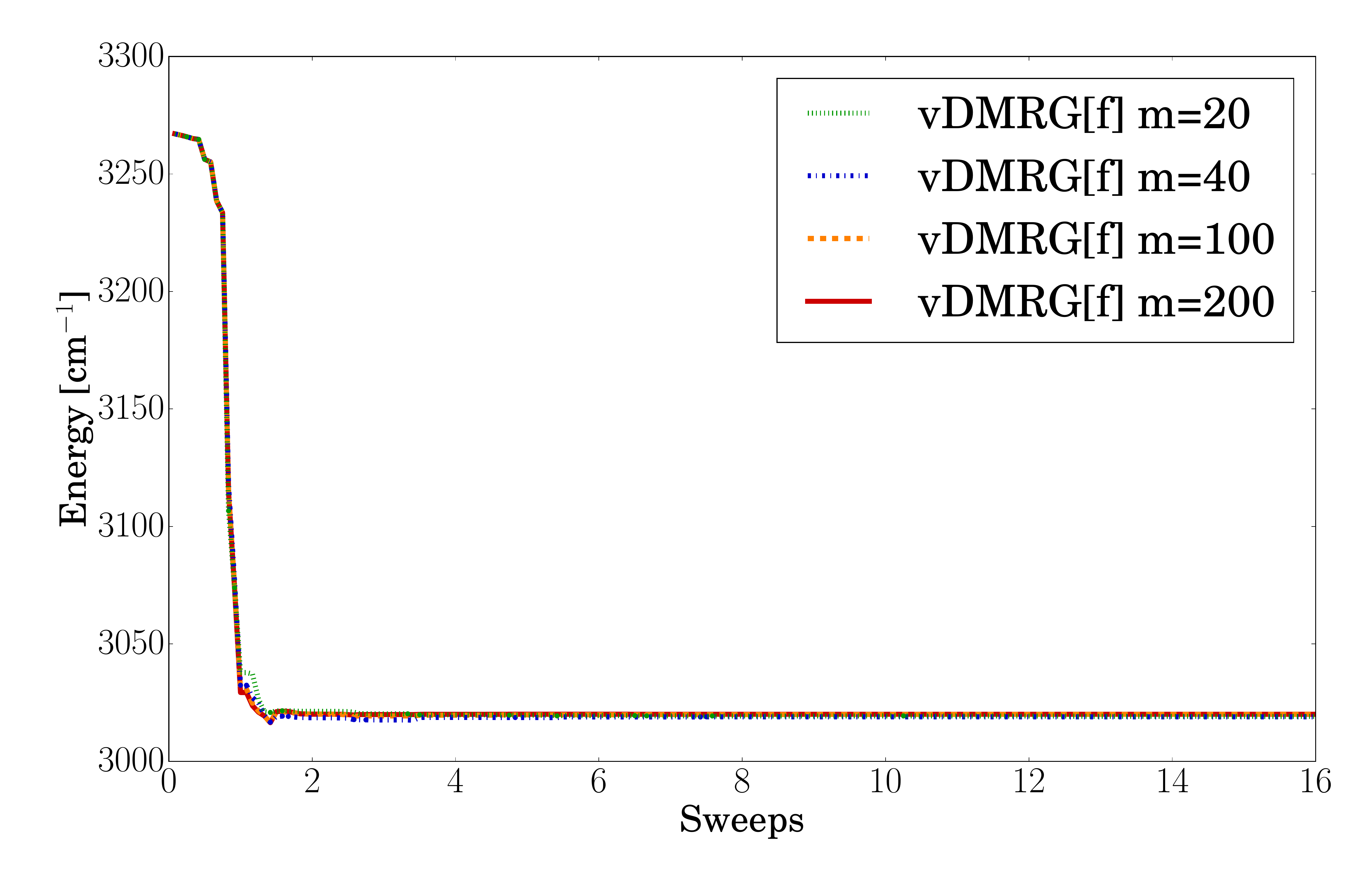}
	\caption{DMRG energies of the $\nu_1$ mode of ethylene as a function of the number of sweeps for different values of the number of renormalized block states $m$. vDMRG[f,MaxO] was employed. The $\omega$ parameter was updated with $\omega_\text{shift}$ = 10 cm$^{-1}$.}
	\label{fig:Conv_CH2_m_Folded}
\end{figure}

The same parameters were applied to calculate the vibrational energies of the second vibration, associated to the $\nu_1$ mode. The vDMRG energy as a function of the sweep number during the MPS optimization with vDMRG[S\&I,MaxO] is reported in Fig.~S4 of Supplementary Material. We note that, for all values of $m$, convergence of the energy is either reached slowly or oscillations are still detected after 10 sweeps. The convergence is, however, much smoother with vDMRG[f,MaxO], as shown in Fig.~\ref{fig:Conv_CH2_m_Folded}. To assess the reliability of the results obtained for the $\nu_1$ mode, we calculated the variance as a function of the number of renormalized block states $m$. The results, reported in Fig.~S5 of Supplementary Material, show that, as expected, the variance decreases monotonically with $m$ and falls below 10~cm$^{-1}$ with $m$=100. We note that the convergence of the variance is much slower than the one of the energy. This suggests that, even though convergence in the energy is reached already with $m$=60, a larger bond dimension is needed to converge the wavefunction as well.

Because of these results, we applied vDMRG[f,MaxO] to the calculation of the vibrational energies of all the 4 C-H stretching modes. The results are reported in Table~\ref{tab:C2H4_CHStretching_Quartic}. As for the low-energy modes, also in this case convergence has already been achieved with $m$=40, and variations below 1 cm$^{-1}$ are observed with higher values of $m$. This suggests that the MPS representation can be efficient also for high-energy excited states.
            
\begin{table*}[ht!]
  \begin{center}
    \begin{tabular}{c|cccc|cc}
    	\hline\hline                                                                                                                                            
    	State        & $m = 20$ & $m = 40$ & $m = 60$ & $m = 100$ & Exp.\cite{Georges1999_C2H4Exp} & Ref.\cite{Delahaye2014_EthylenePES} \\
		\hline             
	    ZPVE         & 11006.2 & 11006.2  & 11006.2  & 11006.1   &                                &                                     \\
		$\nu_{11}$   & 3000.1  &  2979.1  &  2976.8  &  2976.2   &          2988.62               &               2978.87               \\
		$\nu_1$      & 3015.3  &  3018.7  &  3019.7  &  3019.9   &          3021.85               &               3017.05               \\		
		$\nu_5$      & 3067.5  &  3076.1  &  3076.6  &  3076.6   &          3082.36               &               3071.50               \\
		$\nu_9$      & 3098.1  &  3097.9  &  3097.8  &  3097.8   &          3104.87               &               3091.91               \\
		\hline\hline
	\end{tabular}
	\caption{vDMRG[f,MaxO] energies (in cm$^{-1}$) of the 4 C-H stretching modes of ethylene calculated with a shift parameter $\omega_\text{shift}$ of 10 cm$^{-1}$. Calculations were performed with $N_\text{max} = 6$ and a quartic force-field in Cartesian normal coordinates.}
	\label{tab:C2H4_CHStretching_Quartic}
  \end{center}
\end{table*}

\subsection{The protonated sarcosine-glycine dipeptide cation}

We have already studied the sarcosine-glycine dipeptide in its protonated form, SarGly$^+$, in our original work on vDMRG\cite{Baiardi2017_vDMRG} to analyze the scaling of vDMRG for large systems that are difficult to calculate with most state-of-the-art variational approaches. 
However, the standard implementation of vDMRG required some approximations. In fact, instead of the full, fourth-order potential, a reduced dimensionality model, where all the modes below 900 cm$^{-1}$ are treated as harmonic, had to be employed. 
Although this reduced-dimensionality scheme reduced the computational cost of vDMRG, because both the DMRG lattice and the MPO are smaller, the main advantage of this scheme over the full-dimensional treatment is the reduced number of low-energy states.
This simplifies the calculation of vibrational energies with standard vDMRG, because the number of states to be optimized before reaching the fingerprint region is smaller. 
However, with the S\&I variant of vDMRG, this limitation can be overcome, because vibrational excited states are targeted directly, without converging all lower-energy states. If not otherwise specified, the folded variant of vDMRG will be employed, which is, based on the results of the previous section, the most reliable in targeting highly-excited states in regions with a high density of vibrational levels.
Here, the analysis reported in our previous work is improved in two respects: first of all, calculations with a larger Hamiltonian, where all modes under 500 cm$^{-1}$ are treated as harmonic, are reported. 
Hence, the number of normal modes treated as anharmonic increases to 43 compared to 35 in our previous work.\cite{Baiardi2017_vDMRG}
To limit the size of the MPO, only two-mode coupling vibrational constants above 10 cm$^{-1}$ were included in the expansion of the potential energy. Even if this might seem a major approximation, the inclusion of those small terms would increase the size of the MPO without modifying significantly the calculated energies.
Second, higher vibrational levels, where the calculation with standard vDMRG is unfeasible due to the large number of states to orthogonalize to, are studied. 
We recall that, in regions with a high density of states, targeting vibrational excited states with constrained optimizations can be very challenging, due to the high computational cost, but also because of root flipping effects. For this reason, the root-homing algorithm is expected to be instrumental to converge correctly vibrational states of large-size molecules.

The four vibrations of SarGly$^+$ that we study in this work are the two CO, one OH, and one NH stretching modes, all in the high-energy region of the spectrum, for which the advantage of the S\&I variant of vDMRG is maximal. 
A graphical representation of these modes is shown in Fig.~S9 of Supplementary Material.
These vibrations are important from an experimental point of view, since they determine the main features of the bandshape in the fingerprint region of the IR spectra of polypeptides.\cite{Decatur2006_ReviewIRProtein,Barth2007_InfraredProtein} 

In recent work,\cite{Cheng2014_LocalModesVCI} the vibrational energies of four out of these four vibrations were determined with vibrational self-consistent-field calculations (VSCF) for a local-mode Hamiltonian calculated from a B3LYP/6-311+G(d,p) PES. 
For the sake of coherence with this work and with our previous analysis,\cite{Baiardi2017_vDMRG} vDMRG calculations were carried out from a quartic potential in Cartesian normal modes obtained with the same electronic structure model. 

\def\arraystretch{1.5}
\begin{table*}[ht!]
	\begin{center}
		\begin{tabular}{c|ccc|cc}
			\hline\hline
			\multirow{2}{*}{State}  & \multicolumn{3}{c|}{Num. of renormalized states $m$} &
			                          \multirow{2}{*}{Exp.\cite{Johnson2014_SarGlyExp}}    & 
			                          \multirow{2}{*}{Ref.\cite{Cheng2014_LocalModesVCI}} \\
			                 &    20      &    40     &    60       &                  &              \\
			\hline
			   ZPVE          &  36692.8   &  36691.6  &  36691.4    &                  &              \\
			   CO  (1)       &   1729.6   &   1728.6  &   1728.4    &         -        &              \\
			   CO  (2)       &   1784.7   &   1784.1  &   1783.6    &      1788        &    1787      \\
			 NH Amide (3)    &   3300.0   &  3304.5   &  3304.9     &        3370      &    3350      \\
			   OH (4)        &   3485.7   &  3483.6   &  3480.6     &        3570      &    3572      \\
			\hline\hline
		\end{tabular}
		\caption{vDMRG[f,MaxO] energies (in cm$^{-1}$) of the CO, NH and OH stretching modes of SarGly$^+$. The numbers reported in the first column refer to the graphical representation given in Fig.~S8 of the Supplementary Material. Calculations were performed with $N_\text{max} = 6$ and with varying numbers of renormalized states.}
		\label{tab:Dipeptide_frequencies}
	\end{center}
\end{table*}

We begin our analysis from the lowest energy modes among the ones studied here, i.e. the two CO stretching modes. 
The lower-energy mode, referred to as CO(1) in the following, is associated with stretching of the CO bond of the amide group, whereas the one at higher energy, referred to as CO(2) in the following, is associated with the terminal carbonyl group.
CO stretching modes are usually characterized by a low degree of coupling (for this reason, CO stretches are usually studied with reduced-dimensionality schemes,\cite{Kvapilova2015_RDMetals} where only a limited number of modes is treated anharmonically). For this reason, the fully-anharmonic wave function is expected to deviate only slightly from its harmonic counterpart.

The two CO modes are also good examples for probing the reliability of the root-homing variant of vDMRG. In fact, these modes are close in energy, with a separation of 61~cm$^{-1}$ at the harmonic level.
Therefore, the ability of targeting them only with the S\&I formulation of vDMRG would strongly depend on the shift parameter $\omega$. 
However, they correspond to MPSs localized on different parts of the vDMRG lattice, and for this reason the root-homing algorithm allows a clear distinction between them. 

\begin{figure}
	\centering
	\includegraphics[width=.5\textwidth]{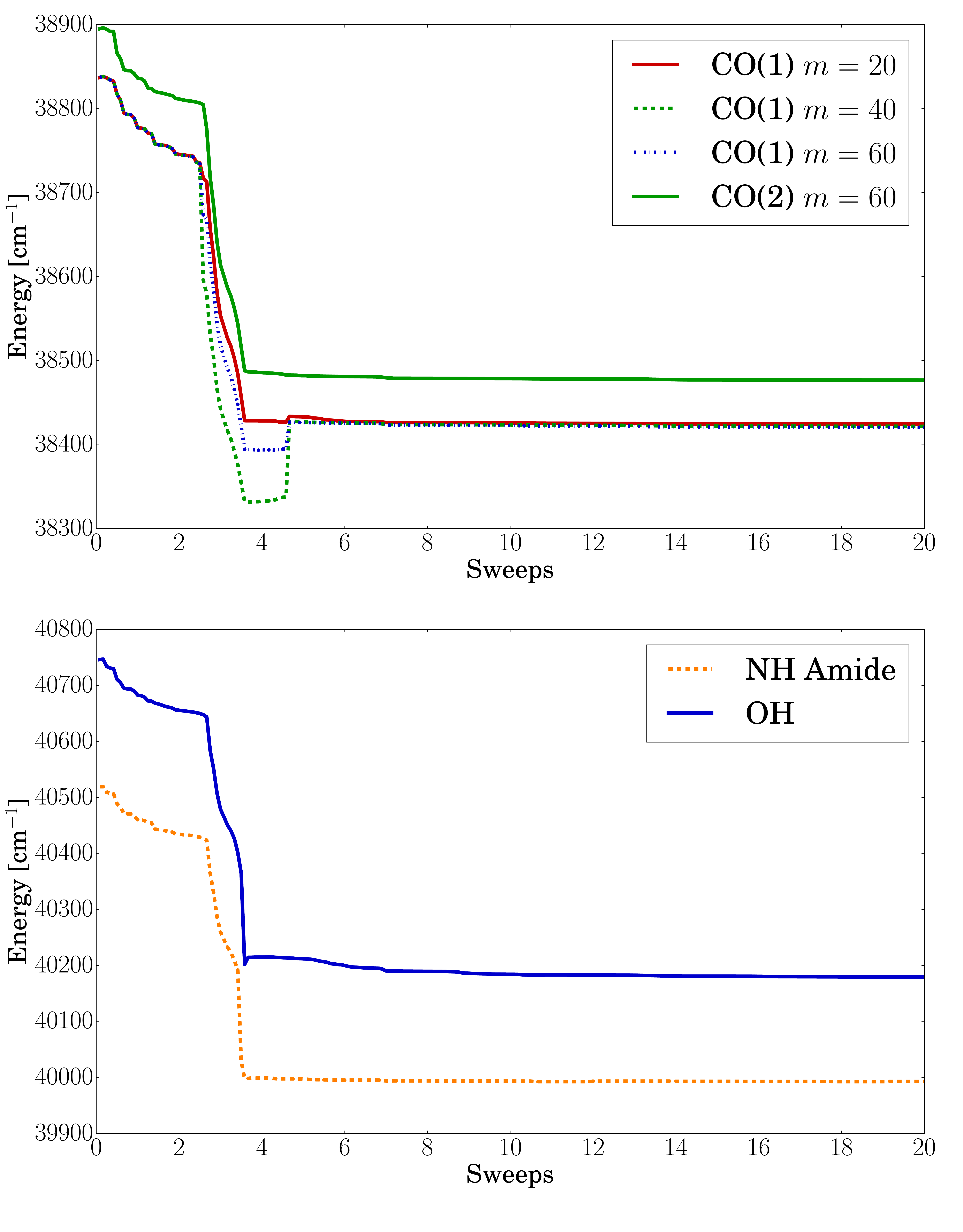}
	\caption{vDMRG[f,MaxO] energy of the CO(1) and CO(2) stretching modes (upper panel) and of the OH and NH amide stretching modes (lower panel) of SarGly$^+$ as a function of the number of sweep in the MPS optimization. Calculations were performed with the vDMRG[f,MaxO] variant, employing different numbers of renormalized block states $m$, and setting $N_\text{max} =6$.}
	\label{fig:CO_SweepConvergence}
\end{figure}

The energy of the first CO mode as a function of the number of sweeps is reported in Fig.~\ref{fig:CO_SweepConvergence} for three different values of $m$ (20, 40, 60 and 100). The figure shows that, in all cases, convergence is achieved within 10 sweeps, as for ethylene. Furthermore, already with $m=40$ renormalized block states, convergence within 1 cm$^{-1}$ is achieved, as highlighted also in Table~\ref{tab:Dipeptide_frequencies}. The same considerations also hold for the second CO stretching mode, as highlighted in the upper panel of Fig.~\ref{fig:CO_SweepConvergence}. It is also worth noting that although, especially in the first sweeps, the energy separation between the two vibrational states is small, the optimization algorithm will follow consistently the correct root.

In the lower panel of Fig.~\ref{fig:CO_SweepConvergence}, the same comparison is reported for the OH and NH amide stretching modes. It is worth noting that these are the two highest fundamental bands of the dipeptide. Therefore, their energy is located in a region with an extremely high density of states. The convergence reported in the lower panel of Fig.~\ref{fig:CO_SweepConvergence} shows that, despite the steep increase in the density of states, also in this case 10 sweeps are sufficient to reach convergence in the energy. This suggests that the accuracy of vDMRG[f] coupled to root-homing is only slightly affected by the density of states. As shown in Table~\ref{tab:Dipeptide_frequencies}, also for these modes convergence with respect to the number of renormalized block states is reached already with $m$=40 states, hence confirming that the MPS parametrization is efficient also for high-lying excited states of large molecules.

\section{Conclusions}

In this work, we extended our previous vDMRG theory\cite{Baiardi2017_vDMRG} to enable a direct targeting and optimization of excited states.
For vibrational calculations, such an approach is mandatory, because the calculation of a vibrational spectrum requires the optimization of a large number of vibrationally excited states.
In a standard DMRG approach, this requires the sequential constrained optimization of all states, which cannot be trivially parallelized. The alternative approach proposed here, which combines a shift-and-invert scheme with root-homing algorithms, makes the calculation of different excited states independent. The theory introduced in this work is, therefore, trivially parallelizable. Multiple calculations associated to different values of $\omega$ can be run in parallel to target simultaneously different regions of the spectrum. The MaxO criterion can be employed to follow specific types of vibrations (e.g, fundamental transitions, overtones or combination bands).

The modular MPS/MPO implementation of \textsc{QCMaquis-V} makes the extensions to the wave function optimization algorithms presented in this work easily applicable to other Hamiltonians, such as the electronic Hamiltonian, for which DMRG has already been well developed in the past twenty years.\cite{whit92,whit93,lege08,chan08,chan09,mart10,mart11,chan11,scho11,kura14,wout14,yana15,szal15,knec16,chan16}

Furthermore, the root-homing algorithms allow us to consistently follow a given vibrational mode during the MPS optimization.
With this feature, we overcome instabilities in the S\&I algorithm and can now target specific prominent bands of molecules with more than 40 vibrational degrees of freedom, as demonstrated at the example of SarGly$^+$.
The detailed analysis of all possible variants of energy-specific vDMRG with different eigensolvers and root-homing procedures allowed us to identify an optimal setup for these calculations that we now define simply as vDMRG without further acronyms.
The harmonic Jacobi--Davidson solver combined with vDMRG[f] and with an overlap-based root-homing (vDMRG[f,MaxO]) in this optimal setup.
Root-homing is realized with an update of the shift $\omega$ in each iteration step, with the shift parameter $\omega_\mathrm{shift}$ = 10~cm$^{-1}$ and a previously defined test MPS with only a single non-vanishing ONV for the calculation of the overlap.
The main limitation of vDMRG[f] is the calculation of the squared value of the vibrational Hamiltonian $\mathcal{H}_\text{vib}^2$, which increases the computational cost with respect to standard vDMRG. For low-lying states, located in regions with a low density of states, the vDMRG[S\&I,MaxO] represents a more efficient alternative, which however does not always converge to the correct root.

The stochastic reconstruction of the VCI determinant space facilitates a simple interpretation of the otherwise rather complicated MPS structure in terms of fundamentals, overtones, and combination bands.

In future work, we will focus on a more flexible representation of the vibrational Hamiltonian. The approach presented here relies on the expansion of the potential in powers of Cartesian normal coordinates. However, for highly anharmonic systems, the harmonic oscillator model does not represent a reliable reference and more refined local basis functions (e.g., the eigenfunctions from VSCF calculations) and functional forms for the potential energy (e.g., expressed in $n$-mode representation) are more suitable. To support such more general representations, different second-quantization forms of the vibrational Hamiltonian, such as the ones proposed in Refs.~\citenum{Christiansen2004_SecondQuantization} and \citenum{Wang2009_SQMCTDH}, must be employed.
For highly anharmonic vibrations, the harmonic wave function might be an inadequate reference for the root-homing algorithm. 
In such cases, preoptimized wave functions with lower bond dimension $m$ or from Hamiltonians with reduced dimensionality would constitute a more reliable reference both as a guess for the MPS optimization and as a test vector for the root-homing.
Other root-homing algorithms, based on different quantities than the overlap, such as transition dipole moments, can be implemented for a more flexible approach.\cite{lube09}
We will furthermore implement the calculation of transition properties between functions expressed as MPSs to provide access to intensities and full anharmonic spectra.

\section*{Supplementary Material}
See supplementary material for additional information on the various optimization algorithms proposed in this work.

\section*{Acknowledgements}

This work was supported by ETH Zurich (ETH Fellowship No. FEL-49 18-1).

\vspace{1cm}
\textbf{References}

%

\end{document}